\documentclass[11pt,letterpaper,superscriptaddres]{article}
\pdfoutput=1
\usepackage{jheppub}
\usepackage[utf8]{inputenc}

\usepackage{color}
\usepackage{amsmath}
\usepackage{mathtools}
\usepackage{esint}
\usepackage{pifont}
\usepackage{bbold}

\usepackage{bbm}
\usepackage{verbatim}   
\usepackage{subfigure}
\usepackage{acronym}

\usepackage{amsfonts}
\usepackage{amssymb}
\usepackage{mathrsfs}
\usepackage{graphicx}
\usepackage{multirow}
\usepackage{slashed}
\usepackage{cleveref}

\usepackage{wasysym}

\usepackage{soul}
\usepackage{cancel}

\usepackage{tensor}

\def\T{\textsf{\tiny T}}
\def\Mat#1{\mathbf{#1}}
\def\mat#1{\mathsf{#1}}

\def\li#1{{\color{BrickRed} #1}}
\def\Li#1{{\color{Cerulean} #1}}
\def\Ri#1{{\color{Blue} #1}}
\def\Hi#1{{\color{OliveGreen} #1}}

\usepackage[dvipsnames]{xcolor}

\preprint{}

\title{Magic Zeroes and Hidden Symmetries}

\author[a,b,c]{Nathaniel Craig,}
\emailAdd{ncraig@physics.ucsb.edu}
\author[d]{Isabel Garcia Garcia,}
\emailAdd{isabel@kitp.ucsb.edu}
\author[e,d]{Arkady Vainshtein,}
\emailAdd{vainshte@umn.edu}
\author[a]{and Zhengkang Zhang}
\emailAdd{zkzhang@ucsb.edu}

\affiliation[a]{Department of Physics, University of California, Santa Barbara, CA 93106, USA}
\affiliation[b]{Physics Division, Lawrence Berkeley National Laboratory, Berkeley, CA 94720, USA}
\affiliation[c]{Berkeley Center for Theoretical Physics, University of California, Berkeley, CA 94720, USA}
\affiliation[d]{Kavli Institute for Theoretical Physics, University of California, Santa Barbara, CA 93106, USA}
\affiliation[e]{FTPI and School of Physics and Astronomy, University of Minnesota, Minneapolis, MN 55455, USA}

\abstract{
	Selection rules arising from accidental or broken symmetries may be sufficiently obscure that their agency is hidden, leading to the appearance of ``magic zeroes'' -- quantities that are suppressed without apparent recourse to a symmetry explanation. Magic zeroes and their corresponding hidden symmetries may shed new light on parametric hierarchies in the Standard Model and beyond. We identify the hidden symmetry responsible for a recently-discovered magic zero, the vanishing of the putative leading contribution to the anomalous dipole moments of the muon upon integrating out weak doublet and singlet vector-like fermions. Some of the tools involved -- spurion analysis leveraging discrete symmetries of the free theory, field redefinitions, spectator fields, and non-supersymmetric non-renormalization theorems -- may prove useful in the hunt for new magic zeroes and their hidden symmetries.
}

\begin{document}
\maketitle
\flushbottom

\section{Introduction}
\label{sec:intro}

Symmetry has long been a guiding principle for understanding the size of parameters both within and beyond the Standard Model (SM). It crucially underlies 't Hooft's notion of technical naturalness \cite{tHooft:1979rat}, namely that suitably normalized parameters may be much smaller than unity only if setting them to zero restores a symmetry of the theory. So great is the success of technical naturalness in explaining most of the observed hierarchies of the SM that it holds sway even over the notable exceptions to the rule. For instance, prevailing approaches to the electroweak hierarchy problem such as supersymmetry or the composite Higgs extend the SM in such a way as to render the Higgs mass technically natural. The lack of evidence (thus far) for these approaches may call into question whether symmetry considerations still have a role to play in solving the puzzles of the SM.

But the role of symmetry in explaining small parameters can often be far from obvious. Symmetries that are badly broken in a full theory may nonetheless give rise to interesting selection rules, particularly when one is only concerned with a subset of the theory -- a given order in loop counting or power counting, for example. Symmetries taken in tandem may control far more than either symmetry in isolation, as manifested by instances of collective symmetry breaking \cite{Arkani-Hamed:2001nha}. Theories may inherit the protection of symmetries from much larger progenitors despite possessing fewer degrees of freedom, as exemplified by the orbifold correspondence \cite{Bershadsky:1998cb} and non-supersymmetric non-renormalization theorems \cite{Alonso:2014rga, Elias-Miro:2014eia, Cheung:2015aba}. Such symmetries may be sufficiently intricate or subtle that their agency is obscured, leading to the appearance of ``magic zeroes'' -- quantities that are suppressed without apparent recourse to technical naturalness.
We will loosely refer to the symmetries underlying these magic zeroes as ``hidden symmetries". These may include symmetries that are fully broken, but are nevertheless responsible for certain cancellations inasmuch as the specific pattern of symmetry breaking controls the form of contributions to symmetry-breaking quantities.
The hidden symmetries underlying these magic zeroes are instances of what we will playfully (and apocryphally) refer to as Wilson's third law, with apologies to Arthur C.~Clarke: 
\begin{quote}
	{\it Any sufficiently advanced symmetry is indistinguishable from magic.}
\end{quote}

Perhaps the surprising lightness of the Higgs is an instance of Wilson's third law. More broadly, perhaps the failure of apparent symmetries to solve the outstanding puzzles of the SM -- the electroweak hierarchy problem, strong CP problem, and cosmological constant problem chief among them -- is an invitation to further explore the space of hidden symmetries. In doing so, it bears remembering that hidden symmetries are discovered as often as they are invented. For every hidden symmetry that has been invented by design, another has been discovered in the course of explaining an observed magic zero. 

The goal of this paper is to look for hidden symmetries that explain magic zeroes. For definiteness, we take as our focal point a specific magic zero, namely the vanishing of the anomalous dipole moments of the muon, associated with a dimension-six operator in the one-loop effective action, for a simple SM extension involving heavy vector-like fermions that are $SU(2)_L^{}$ singlets and doublets. 
In Ref.~\cite{Arkani-Hamed:2021xlp} this was attributed to a ``total derivative phenomenon,'' in lieu of an apparent symmetry at play. In this paper we show that the magic zero can be explained by hidden symmetries.\footnote{The fact that the magic zero of Ref.~\cite{Arkani-Hamed:2021xlp} can be explained by symmetries does not lessen the significance of total derivative phenomena, which illuminate a number of interesting relations between operator coefficients that would otherwise remain obscure.} Along the way, we observe some rules of thumb that may prove more broadly useful in the search for hidden symmetries. For example, it is helpful to consider the full symmetries of the free theory (including those that are often neglected in analyzing fermionic theories); in this case a nontrivial discrete subgroup of these symmetries plays a particularly valuable role. The relevance of this discrete symmetry to the magic zero is perhaps most apparent in the mass basis after electroweak symmetry breaking, an appropriate arena for addressing questions of Wilsonian naturalness. The relevance of the discrete symmetry in the unbroken phase is less obvious, but can be made apparent by a judicious choice of field redefinitions. Such redefinitions run against the familiar grain, leveraging vector-like fermion mass terms to eliminate marginal non-derivative interactions in favor of irrelevant derivative ones (similar to the case of axion EFT below the Peccei-Quinn breaking scale~\cite{Georgi:1986df}).
It is also useful, though not obligatory, to view the model as a subset of a larger theory, in which the additional states are mere spectators to the physics of interest. Finally, a spurion analysis organized around internal symmetries can be usefully supplemented by non-supersymmetric non-renormalization theorems based on helicity selection rules. Some or all of these features may prove helpful in identifying or engineering new magic zeroes.

This paper is organized as follows: In \cref{sec:puzzle} we first review the global symmetries of a free, massless Dirac fermion before turning to the model of Ref.~\cite{Arkani-Hamed:2021xlp}, where we repeat the one-loop calculation of the muon anomalous dipole moments. These yield a ``magic zero'' when the Higgs is taken to be massless: in the unbroken phase it manifests as a vanishing coefficient for the appropriate dimension-6 dipole operator, while in the broken phase it corresponds to a vanishing dipole moment at leading order in $v$. We begin the search for a symmetry explanation of the magic zero in \cref{sec:symmetry}, setting up a spurion analysis in a toy model of scalars, massive fermions, and massless fermions that controls the form of possible contributions to the dipole moments of the massless fermions. Mapping the full model in the mass eigenbasis onto this toy model, we show that symmetries of a subset of the theory force the dipole moments to vanish at leading order in $v$. This naturally invites an analogous explanation in the unbroken phase, which we present in \cref{sec:unbroken}. Here the constraints imposed by symmetry are made most apparent by carrying out a field redefinition, and force the corresponding dimension-6 operator to vanish. In \cref{sec:conspiracy} we turn to the pragmatic question of how the vanishing dipole would appear to an effective field theorist living at intermediate energies, and the extent to which they might construe the magic zero as the result of a UV-IR conspiracy. In \cref{sec:conclusions} we distill some general lessons in the hunt for UV-IR conspiracies and symmetry explanations of magic zeroes.

\section{The magic zero}
\label{sec:puzzle}

An apparent ``magic zero" was recently discussed in Ref.~\cite{Arkani-Hamed:2021xlp}, in the context of what is otherwise a remarkably simple model. The theory considered in Ref.~\cite{Arkani-Hamed:2021xlp} (and studied previously in \cite{Kannike:2011ng, Freitas:2014pua}) extends the SM by a pair of vector-like $SU(2)_L^{}$-doublets, as well as a pair of SM singlets. At one loop, charged SM leptons acquire a mass correction in this model, yet the leading contribution to their electromagnetic dipole moments is, somewhat surprisingly, absent.

The purpose of this section is to summarize the puzzle raised in Ref.~\cite{Arkani-Hamed:2021xlp} and provide a complementary perspective on some aspects of the result. We begin with a brief review of the different symmetry structure of the mass and dipole operators of a Dirac fermion in \cref{sec:U(2)}. In \cref{sec:1loop}, we introduce the model that is the focus of Ref.~\cite{Arkani-Hamed:2021xlp}, and discuss (and partly clarify) the relevant aspects of the one-loop calculation. In \cref{sec:infrared} we emphasize the role played by the infrared dominance of the loop diagram, which may provide helpful guidance in the search for analogous phenomena elsewhere. 

\subsection{Symmetries of a Dirac fermion}
\label{sec:U(2)}

The Lagrangian of a massless Dirac fermion reads
\begin{equation}
\mathcal{L}_\text{free} = \bar \Psi i \gamma^\mu \partial_\mu \Psi = e^\dagger i \bar \sigma^\mu \partial_\mu e + {e^c}^\dagger i \bar \sigma^\mu \partial_\mu e^c ,
\end{equation}
where in the last step we have written the Lagrangian in terms of left-handed two-component Weyl fermions $e$ and $e^{c}$ after substituting $\Psi = (e, {e^c}^\dagger)^T$. This expression can be further written as
\begin{equation}
\mathcal{L}_\text{free} = \chi^\dagger_{i} i \bar \sigma^\mu \partial_\mu \chi^{i} , \qquad \text{where} 
\qquad \chi^{1} \equiv e, ~~ \chi^{2}  \equiv e^{c }. 
\end{equation}
This makes it manifest that the Lagrangian of a massless Dirac fermion is invariant under a $U(2)$ global symmetry under which $\chi$ transforms as a doublet. We will write this as $U(2) = U(1)_A^{} \times SU(2)_V^{}$, with the $U(1)_V^{}$ subgroup generated by the $\tau^3$ generator of $SU(2)_V^{}$.

In two-component notation, which we will use throughout this work, Dirac mass and electromagnetic dipole operators take the form
\begin{equation}\label{Eq:massanddipole1}
\left( - m \epsilon^{\alpha \beta} e_\beta e^c_\alpha + \tau\, F^{\alpha \beta} e_\beta e^c_\alpha\right) + \text{h.c.} ,
\end{equation}
where we have made Lorentz indices explicit, and 
$F^{\alpha\beta} = -\epsilon^{\alpha\gamma}\,(\sigma^{\mu\nu})\indices{_\gamma^\beta}F_{\mu\nu}$
is symmetric under $\alpha \leftrightarrow \beta$.\footnote{We follow the conventions of Ref.~\cite{Dreiner:2008tw} where $\sigma^{\mu\nu}$ is defined with lower-upper spinor indices: $(\sigma^{\mu\nu})\indices{_\alpha^\beta}\equiv \frac{i}{4}\,\bigl( \sigma^\mu_{\alpha\dot\gamma}\, \bar\sigma^{\nu\dot\gamma\beta} -\sigma^\nu_{\alpha\dot\gamma}\, \bar\sigma^{\mu\dot\gamma\beta}\bigr)$. In this convention, $\sigma^{i0} = -\sigma^{0i} = \frac{i}{2}\vec{\sigma}^i$ and therefore $\sigma^{\mu\nu}F_{\mu\nu} = -\vec{\sigma} \cdot(\vec B + i \vec E) $.}
The real part of the coefficient $\tau$ is the anomalous magnetic moment of the fermion, while its imaginary part gives the electric dipole moment, {\it i.e.}\
\begin{equation} \label{Eq:magelecdipole}
\tau=d_{m} +i d_{e} \,.
\end{equation}

In terms of the left-handed Weyl doublet, $\chi^{i}$, the expression in \cref{Eq:massanddipole1} reads
\begin{equation} \label{Eq:massanddipole}
\frac{1}{2} \left( - m_{ij}^{} \epsilon^{\alpha \beta} \chi^i_\beta \chi^j_\alpha - \tau \,\epsilon_{ij}^{} F^{\alpha \beta} \chi^i_\beta \chi^j_\alpha \right) + \text{h.c.} ,
\end{equation}
with the symmetric mass matrix $m_{ij}^{}$, $m_{12}^{}= m_{21}^{} = m$. A mass term breaks the global $U(2)$ symmetry all the way down to the $U(1)_V^{}$ factor.\footnote{\,It means we are dealing with the Dirac mass. Adding nonvanishing $m_{11}^{},~m_{22}^{}$ to the mass matrix $m_{ik}^{}$ would lead to the Majorana masses and break $U(1)_V^{}$.}
On the other hand, since $\epsilon_{ij}$ is the $SU(2)$-invariant tensor, a non-zero dipole leaves unbroken the \emph{entire} $SU(2)_V$ subgroup. Note that \cref{Eq:massanddipole} makes apparent a discrete subgroup of $U(2)$ useful for distinguishing masses and dipoles: under the $\mathbb{Z}_2$  that exchanges $\chi^1 \leftrightarrow \chi^2$, the mass is even ($m_{12}^{} = m_{21}^{}$) while the dipole is odd ($\tau \epsilon_{12}^{}=-\tau \epsilon_{21}^{}$). This subgroup is just charge conjugation restricted to the fermion sector.

As a result, radiative corrections generating a dipole operator do not necessarily imply a non-zero mass. This observation was made and exploited by Voloshin in Ref.~\cite{Voloshin:1987qy} to construct models featuring a neutrino magnetic dipole moment at one-loop, but vanishing neutrino mass. What naively would have been an ``unnaturally" small mass-to-dipole ratio is achieved in Ref.~\cite{Voloshin:1987qy} by extending the SM with an additional sector that preserves the $SU(2)_V^{}$ symmetry of the neutrino sector. 

The opposite situation -- a small dipole-to-mass ratio -- is clearly more sinister. Radiative corrections generating a fermion mass indicate that the $U(1)_A^{}$ factor is broken, and therefore there is a priori no symmetry preventing the generation of an electromagnetic dipole. This is the situation that arises in the model of Ref.~\cite{Arkani-Hamed:2021xlp}, as we now discuss.

\subsection{Mass vs.\ dipole at one loop}
\label{sec:1loop}

The model that is the focus of Ref.~\cite{Arkani-Hamed:2021xlp} extends the SM by a pair of vector-like $SU(2)_L^{}$-doublets $L^c=(L^{c+}, L^{c\,0})^T$ and $L=(L^0, L^-)^T$, carrying hypercharge $\pm \frac{1}{2}$, as well as two SM-singlets $S^c$ and $S$. We  follow here the notation of  Ref.~\cite{Arkani-Hamed:2021xlp} except that we use a convention where the Higgs doublet carries hypercharge $+\frac{1}{2}$. 
We add to the SM the following mass and Yukawa terms:
\begin{equation}
\mathcal{L}_{\text{add}} = \left\{ - m_L L^c L - m_S S^c S - Y'_V H L S^c - Y_V H^\dagger L^c S - Y_L H l S^c - Y_R H^\dagger L e^c \right\} + \text{h.c.}
\label{eq:model}
\end{equation}
The above interactions break the various axial and vector $U(1)$ symmetries of the free fermion theory, with the exception of the diagonal $U(1)_V^{}$ factor corresponding to fermion number. Non-zero vector-like masses, $m_L$ and $m_S$, break the $U(1)_A^{}$ symmetries acting on the doublet and singlet pairs of fermions, whereas fully breaking the axial symmetry of the SM leptons requires $Y_L^{}, Y_R^{} \neq 0$ as well as either $Y_V^{}$ or $Y'_V$ to be non-vanishing. As a result, any contribution to either the mass or dipole of the SM fermions must be proportional to a suitable combination of all three Yukawas.

Before moving on, let us note that, as in the SM, the space-time symmetries $P$ and $C$ are broken in this model. $T$ is also broken in general: of the (a priori) six complex parameters introduced in \cref{eq:model} one phase remains physical.\footnote{\, For simplicity, and without loss of generality, we take this phase to live in $Y'_V$ in this section. In subsequent sections we will find it useful to remain fully general, treating all the different parameters as complex.}
Throughout this work, we will assume a hierarchy $m_L, m_S \gg v$, and treat the SM leptons as massless. Also we omit the gauge field part of SM besides the electromagnetic external field which probes dipole moments. Loops are formed due to Higgs field interactions, with charged Higgs components appearing as internal lines. The same results can be obtained in the unitary gauge in the limit of vanishing gauge coupling $g \to 0$ where only
longitudinal $W$'s propagate in the loop.

\begin{figure}[h]
	\centering
	\includegraphics[scale=1]{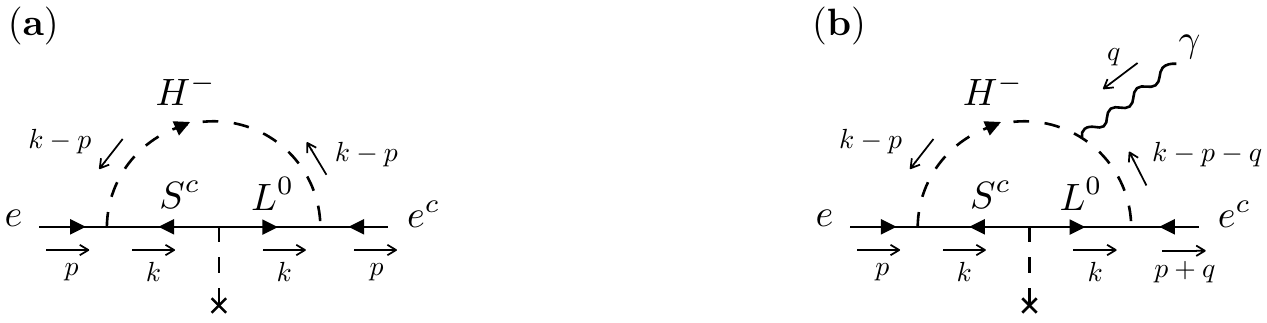}
	\caption{One-loop diagrams providing the (a priori) leading contribution to {\bf (a)} the mass, and {\bf (b)} electromagnetic dipole moments of charged SM leptons in the model of \cref{eq:model} with $Y_V^{}=0$. Both diagrams are proportional to the combination of couplings $Y_L^{} Y_R^{} Y_V^{\prime*}$. }
	\label{fig:oneloop_2}
\end{figure}

Integrating out the vector-like fermions at tree level does not generate either a mass or a dipole moment for the charged SM fermions. Instead, the leading contributions arise at one-loop, and, as anticipated, are linear in $Y_V$ or $Y'_V$. We will focus on contributions involving $Y'_V$ in the rest of this section, and relegate the discussion of those proportional to $Y_V$ to \cref{sec:app_calculation}.
\Cref{fig:oneloop_2} shows the (a priori) leading one-loop diagrams involving $Y'_V$. Indeed, upon setting the Higgs to its vacuum expectation value, the diagram of \cref{fig:oneloop_2}(a) leads to a contribution to the lepton mass of the form
\begin{equation} \label{eq:deltammu}
\delta m (\mu) = 
\frac{v}{\sqrt{2}} \frac{Y_L Y_R Y'^*_V}{16 \pi^2 (m_L^2 - m_S^2)} \left( m_L^2 \log \frac{\mu^2}{m_L^2} - m_S^2 \log \frac{\mu^2}{m_S^2} \right) ,
\end{equation}
where $\mu$ denotes the renormalization scale (or a UV cut-off in a mass-dependent scheme).

However, the contribution to the electromagnetic dipole from the diagram of Fig.~\ref{fig:oneloop_2}(b) vanishes, as it was observed in Ref.~\cite{Arkani-Hamed:2021xlp}.
It is not hard to verify this is the case.
Setting aside overall multiplicative vertex factors, the corresponding amplitude is proportional to
\begin{align} \label{eq:Mdiople}
i \mathcal{M}^\mu 	& \propto \int \frac{d^4 k}{(2 \pi)^4} \frac{k^2 (2 k^\mu - 2 p^\mu - q^\mu)}{(k-p)^2 (k-p-q)^2 (k^2 - m_L^2) (k^2 - m_S^2)} \bar u (p+q) P_L u(p) ,
\end{align}
where we use the four-component fermions $u$ and the chiral projection $P_{L}=(1-\gamma^{5})/2$. Note that the $i \epsilon$ terms in the propagators are not shown having in mind Wick rotation. Expanding this expression to linear order in external momenta, one finds
\begin{align} \label{eq:Mmu_linear}
i \mathcal{M}^\mu 	& \propto \int \frac{d^4 k}{(2 \pi)^4} \frac{ \bar u (p+q) P_L u(p)}{k^2 (k^2 - m_L^2) (k^2 - m_S^2)} \left\{ - (2 p^\mu + q^\mu) + 2 k^\mu \left( 1 + \frac{2 k \cdot (2 p + q)}{k^2} \right) \right\} .
\end{align}
The term linear in $k$ can be ignored, as it vanishes upon integration. Making the replacement $4 k^\mu k^\nu \rightarrow k^2 \eta^{\mu \nu}$ in the last term of the previous expression\footnote{\,We set $d=4$ as the integral in question is finite; no regularization is required in this example.} we find
\begin{align}
i \mathcal{M}^\mu \propto \int \frac{d^4 k}{(2 \pi)^4} \frac{ \bar u (p+q) P_L u(p)}{k^2 (k^2 - m_L^2) (k^2 - m_S^2)} \left\{ - (2 p^\mu + q^\mu) + (2 p^\mu + q^\mu) \right\} = 0 .
\label{eq:amplitude_zero}
\end{align}
Since the integrand vanishes at linear order in $p$ and $q$, there is indeed no contribution to the electromagnetic dipole term from this diagram.

In Ref.~\cite{Arkani-Hamed:2021xlp} the vanishing of the diagram in \cref{fig:oneloop_2}(b) is ascribed to the integrand of the loop function being a total derivative that vanishes in the limits $|k_E| \rightarrow 0$ and $|k_E| \rightarrow \infty$. This justification may seem different from the one we have presented here, where we have seen that the integrand itself is zero at linear order in external momenta. The difference between our presentation and that in Ref.~\cite{Arkani-Hamed:2021xlp} stems from a relabelling of the loop momenta in \cref{fig:oneloop_2}(b). Shifting $k \rightarrow k + p + q$ modifies the loop integrand by a total derivative term that vanishes upon integration. The physical significance of this total derivative is unclear given that it can be removed by an unphysical relabelling of internal momenta, motivating the search for a symmetry-based explanation.\footnote{
	One may take advantage of this relabelling freedom to show that certain classes of higher-dimensional operators are related by writing loop integrands as total derivatives, as discussed in Ref.~\cite{Arkani-Hamed:2021xlp}. As long as fermion masses are the only scales appearing in the loop, these relations also admit a simple symmetry explanation: first, the fact that the two classes of operators arise from diagrams with identical fermion lines means that operator coefficients at the same mass dimension ($\mathbf{B}$ and $\mathbf{\tilde B}$, $\mathbf{C}$ and $\mathbf{\tilde C}$, etc.\ in the notation of Ref.~\cite{Arkani-Hamed:2021xlp}) must have identical flavor structures and therefore be proportional to each other up to logarithms; meanwhile, logarithms can only arise from divergences of sub-diagrams, which again must agree between the two sets of diagrams. Note that $\mathbf{B}\propto\mathbf{\tilde B}$ and $\mathbf{B}=0$ do not necessarily indicate that $\mathbf{\tilde B}$ (corresponding to the dipole) vanishes because the symmetry argument above does not exclude a zero proportionality constant --- we will need slightly more complicated symmetry analyses to show that the dipole indeed vanishes, as discussed in the rest of the paper. The symmetry argument also fails if there are additional scales appearing in the loop, {\it e.g.}\ from massive scalars. We thank Nima Arkani-Hamed for discussion of this point.
}

Also, as noted in Ref.~\cite{Arkani-Hamed:2021xlp}, the vanishing of the electromagnetic dipole moments does not survive beyond the leading analysis summarized here. Non-zero contributions are present at two-loop order, and also at one-loop order proportional to $m_W^2$. As our focus here is on finding a symmetry-based explanation for the absence of what would have been the leading dipole term, we will not be concerned with these non-zero pieces. However, let us note that had we included a non-zero mass for the charged scalar in the corresponding propagators appearing in \cref{eq:Mdiople}, the cancellation in \cref{eq:amplitude_zero} would not have taken place.

\subsection{Infrared dominance}
\label{sec:infrared}

Before setting off in search of a symmetry explanation for the magic zero, it is useful to consider one last derivation that still makes use of the properties of the loop integrand by connecting the one-loop dipole to the infrared limit of certain scattering amplitudes. This argument (which is closely related to arguments in Ref.~\cite{Arkani-Hamed:2021xlp}) is fairly general and informs the discussion of apparent UV-IR conspiracies in \cref{sec:conspiracy}. 

Consider the general scenario of a charged massless Dirac fermion $\Psi = (e, {e^c}^\dagger)^T$ interacting with an oppositely-charged massless scalar field $\phi$ and any number of massive, neutral fermions. We can write the one-loop diagram contributing to the dipole as
\begin{align} \label{eq:Mdiople1}
i \mathcal{M}^\mu = e\, (e^{c}_{\alpha} e^{\alpha})
\int \frac{d^4 k}{(2 \pi)^4} \frac{2 k^\mu - 2 p^\mu - q^\mu}{((k-p)^2-\lambda^{2}) ((k-p-q)^{2} -\lambda^{2})}\times h(k^{2}) ,
\end{align}
where $e$ refers to the electric charge, and $(e^{\alpha}e^{c}_{\alpha})h(k^{2})$ represents the the scattering process $\phi(k-p) + e(p) \to \phi (k-p-q) + \overline{e^{c}}(p+q)$.
We have also introduced a small mass $\lambda$ for the scalar $\phi$ as an infrared regulator, with the intention of taking the limit $\lambda \to 0$ at the end. We further assume UV convergence (i.e.~the scattering amplitude $h(k^{2})$ goes to zero at large $k^{2}$), which is certainly provided within renormalizable theories. 

For the purposes of extracting dipole moments we can limit ourselves to a first order expansion in the external momenta $p$ and $q$. Working at this order and substituting $4k^{\mu}k^{\nu}\to k^{2}\eta^{\mu\nu}$ (possible thanks to the IR regulator), we arrive at
\begin{align} \label{eq:Mmu_linear2}
i \mathcal{M}^\mu= e\, (e^{c}_{\alpha} e^{\alpha})(2 p^\mu + q^\mu)\int \frac{d^4 k}{(2 \pi)^4} \frac{\lambda^{2}}{(k^2-\lambda^{2})^{3}}  \times h(k^{2}) .
\end{align}
After Wick rotation the integration over $k$ -- clearly dominated by $k\sim \lambda$ -- is easily done, and in the $\lambda \to 0$ limit we obtain for the dipole amplitude
\begin{align} \label{eq:Mmu_linear3}
\mathcal{M}^\mu=- \frac{e}{32\pi^{2}}\,h(0) (e^{\alpha}e^{c}_{\alpha})(2 p^\mu + q^\mu) = 
 \frac{ie}{16\pi^{2}}\,h(0) \,e^{c}\sigma^{\mu\nu}q_{\nu}e,
\end{align}
where we used $e^{c}D^{2}e=i e^{c}F_{\mu\nu}\sigma^{\mu\nu}e$ for an on-shell massless fermion.
The anomalous dipole moments are then
\begin{equation} \label{Eq:magelecdipole1}
\tau=d_{m} +i d_{e}= \frac{eh(0)}{16\pi^{2}}\,.
\end{equation}

The contribution to the dipole is governed by the far infrared, and the final form of the dipole is controlled by the infrared behavior of the $\phi + e \rightarrow \phi + \overline{e^c}$ amplitude. 
In Ref.~\cite{Arkani-Hamed:2021xlp}, the same result is derived without introducing a nonzero $\lambda$. 
In that case, the replacement $4k^{\mu}k^{\nu}\to k^{2}\eta^{\mu\nu}$ cannot be made because of IR divergence; instead, the integrand is shown to be a total derivative, which upon integration picks up an IR boundary term. This underlines the advantage to framing the integrand as a total derivative, in that the IR dominance is apparent without regularization. 
Our derivation, with the use of an IR regulator $\lambda$, does not resort to the total derivative phenomenon but also underlines the IR dominance of the dipole.

The tree-level scattering amplitude $\phi + e \rightarrow \phi + \overline{e^c}$ that controls the form of the dipole is specified by the neutral fermions exchanged. The simplest case is to have just one massive Dirac fermion, consisting of two left-handed spinors $\psi$ and $\psi^{c}$ with Lagrangian
\begin{align}
\label{eq:toy1}
\Delta\mathcal{L}	 = \big\{- m \, \psi^c \psi + y_L^{} \, \phi (\psi^c e) + y_R^* \, \phi^* (e^c \psi) \big\} + \text{h.c.} 
\end{align}
The amplitude $h(k^{2})$ in this case is 
\begin{align}
\label{eq:toy2}
h(k^{2})=\frac{y_{L}y_{R}^{*}\,m}{k^{2}-m^{2}}\,, \qquad h(0)=-\frac{y_{L}y_{R}^{*}}{m}\,.
\end{align}
The anomalous dipole moments in \cref{Eq:magelecdipole1} are then given by the value of $h(0)$.

More generally there can be some number of massive Dirac fermions. The magic zero in the model of Ref.~\cite{Arkani-Hamed:2021xlp} arises when
the sum of $y_{L}y_{R}^{*}/m$ over two Dirac fermions vanishes. The authors of \cite{Arkani-Hamed:2021xlp} provided a clever explanation of the vanishing of $h(0)$ using the vanishing of the operator $(l H^{\dagger}) (HH) e^{c}$, where $l$ and $H$ are lepton and Higgs doublets. As we will see in \cref{sec:symmetry}, symmetries can be used to understand why the sum of $y_{L}y_{R}^{*}/m$ over two Dirac fermions vanishes. At face value, symmetry provides the more powerful explanation of the magic zero, in that it both controls the form of the loop amplitude leading to infrared dominance {\it and} orchestrates cancellations among different infrared contributions. 

Ultimately, this motivates pursuing a complete symmetry explanation, in both the mass basis (\cref{sec:symmetry}) and the gauge basis (\cref{sec:unbroken}). But the intermediate level of explanation based on infrared dominance (or total derivative phenomena) is useful insofar as it helps to understand the magic zero starting from the form of the integrand. Infrared dominance further helps to understand the emergence of apparent UV-IR conspiracies, as we will see in \cref{sec:conspiracy}.

\section{Symmetries at work: mass basis}
\label{sec:symmetry}

We now turn our attention to whether the vanishing of (what would have been) the leading one-loop contribution to the electromagnetic dipole moment can be understood in terms of symmetries. The fact that the absence of a dipole term only holds up to a certain order indicates that the symmetry we are after cannot be a symmetry of the full model. Instead, if symmetry is indeed at play, it will likely only involve the field content and interactions that partake in radiative corrections at the relevant order in a perturbative expansion.

In the remainder of this section, we show how the magic zero reviewed in \cref{sec:puzzle} can be understood in terms of symmetries in the mass basis (electroweak broken phase) calculation. The magic zero can also be explained by a gauge basis (electroweak unbroken phase) analysis, as we will discuss in the next section. 
For simplicity, we focus on the case where $Y_V = 0$ in this section. This allows us to simplify the technicalities of our discussion without losing any of the qualitative insights. We relegate the discussion of the case with non-zero $Y_V$ to \cref{sec:app_broken}.

\subsection{Toy model}
\label{sec:toy}

Let us first consider a simplified toy model containing two electrically neutral massive Dirac fermions, with left-handed Weyl components $(\psi, \psi^c)$ and $(\hat\psi, \hat\psi^c)$, respectively, a complex bosonic field $\phi$, describing a charged scalar, as well as a massless charged Dirac fermion with Weyl components $(e, e^c)$. Aside from the appropriate kinetic terms, which include electromagnetic interactions via covariant derivatives, our toy model features the following mass and interaction terms:
\begin{align}
\Delta\mathcal{L}	 =&\; \Big\{ - m \, \psi^c \psi + y_L^{} \, \phi (\psi^c e) + y_R^* \, \phi^* (e^c \psi) \Big\} + \text{h.c.} 
\nonumber\\
& + \Big\{ - \hat m \, {\hat \psi}^c \hat \psi + \hat y_L^{} \, \phi ({\hat \psi}^c e) + {\hat y}_R^* \, \phi^* (e^c \hat \psi) \Big\} + \text{h.c.} 
\label{eq:toy}
\end{align}
The above terms break the various symmetries of the free theory. In the fermion sector, the $U(6)$ global symmetry of the free Lagrangian is broken down to a single $U(1)_V^{}$ factor, corresponding to fermion number.

We can ``restore" the symmetry enjoyed by the free theory by treating the various mass and Yukawa couplings as spurions transforming non-trivially under it. Only the $U(1)^6 \times P_\psi^{}$ subgroup of the full symmetry of the fermion sector will be relevant to our discussion. The $U(1)$ factors correspond to the various axial and vector symmetries of the three pairs of Weyl fermions, whereas $P_\psi^{}$ is a $\mathbb{Z}_2$ discrete permutation symmetry exchanging $\psi \leftrightarrow \hat\psi$, $\psi^c \leftrightarrow \hat\psi^c$, and which acts ``spuriously" on the model parameters as $x \leftrightarrow \hat x$ for $x=m, y_L^{}, y_R^{}$.

We are now ready to identify the form of the coefficients that can be generated for both mass and dipole operators, see \cref{Eq:massanddipole1}. In both cases, they must carry two units of the $U(1)_A^{}$ charge corresponding to the $(e, e^c)$ pair, and be even under all the other abelian factors, including the discrete factor $P_\psi$.
At mass dimension one, the unique object carrying the appropriate $U(1)$ charges that is invariant under $P_\psi$ is given by
\begin{equation}
\delta m =  
y_L^{} y_R^* \,m^* \left( \alpha\log \frac{\mu^2} {|m|^2}+\beta\right) + 
{\hat y}_L^{} {\hat y}_R^* \,{\hat m}^* \left(\alpha\log \frac{\mu^2}{|{{\hat m}}|^2} +\beta\right)
,
\label{eq:delta_m}
\end{equation}
where $\alpha$ and $\beta$ are numerical coefficients that cannot be obtained within a spurion analysis (explicit computation reveals that $\alpha=\frac{1}{16\pi^2}$ and $\beta=0$). Logarithms will generally be present, as renormalizability of the model does not preclude the presence of a counter-term.

The coefficient of dipole operator, $\tau$, on the other hand, has mass dimension $-1$. In this case, the unique object consistent with all the spurion charge assignments reads:
\begin{equation}
	\delta \tau = \gamma \, e \left\{ \frac{y_L^{} y_R^*}{m} + \frac{{\hat y}_L^{} {\hat y}_R^*}{\hat m}  \right\} ,
\label{eq:delta_d}
\end{equation}
reproducing the result in \cref{sec:infrared}.
Notice the absence of logarithms in \cref{eq:delta_d}.  Logarithms here would necessarily involve the renormalization scale $\mu$, pointing to a divergence. 
As a higher-dimensional operator generated in a renormalizable theory, the dipole must be UV-finite. Meanwhile, the absence of IR divergences may be seen most clearly from the perspective of the low-energy theory, where the tree-level dimension-5 irrelevant operators generated by integrating out the heavy fermions are incapable of renormalizing the dipole operator at one loop. This may be understood from one-loop non-supersymmetric non-renormalization theorems based on super-operator embedding \cite{Elias-Miro:2014eia} or helicity selection rules \cite{Cheung:2015aba}.

So far, our discussion has proceeded under the assumption that the charged scalar is massless. Although it won't be relevant to our later discussion, it is worth mentioning how our previous analysis is modified when $m_\phi^2 \neq 0$. Assuming $m_\phi^2 \ll m^2, \hat m^2$, we can build additional objects with the appropriate $U(1)$ charge assignments as an expansion in the scalar mass-squared parameter. For example, at first order in $m_\phi^2$, the following combination provides an additional ``legal" contribution to the coefficient of a dipole term:
\begin{equation}
	\delta \tau^{(1)} =  e \left\{ \frac{y_L^{} y_R^*}{m} \frac{m_\phi^2}{|m|^2} \left(\alpha'\log \frac{|m|^2}{m_\phi^2} +\beta'\right) + \frac{{\hat y}_L^{} {\hat y}_R^*}{\hat m} \frac{m_\phi^2}{|\hat m|^2} \left(\alpha'\log \frac{|\hat m|^2}{m_\phi^2}+\beta' \right) \right\} .
\label{eq:delta_d1}
\end{equation}
Notice that, unlike in \cref{eq:delta_d}, a non-zero $m_\phi^2$ provides an additional scale that allows for the presence of renormalization scale-independent logarithmic factors.

\begin{figure}[h]
	\centering
	\includegraphics[scale=1]{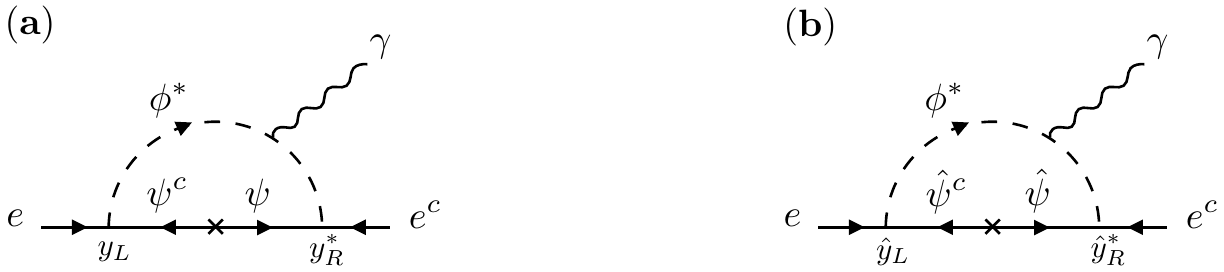}
	\caption{One-loop diagrams contributing to the electromagnetic dipole moments of a charged Dirac fermion in the model of \cref{eq:toy}. {\bf (a)} and {\bf (b)} feature a Dirac fermion with mass $m$ and $\hat m$ respectively propagating inside the loop. For $m_\phi^2 = 0$, the amplitude is of the form given in \cref{eq:delta_d}.}
	\label{fig:oneloop_dipole}
\end{figure}

Diagramatically, there are two graphs contributing to the mass and dipole moments of the massless fermion, as shown in \cref{fig:oneloop_dipole}. Explicit calculation reveals that the contribution to both the mass and dipole coefficients at zeroth order in the scalar mass parameter are indeed of the form given in \cref{eq:delta_m,eq:delta_d}, whereas the form of \cref{eq:delta_d1} captures the leading correction to the dipole coefficient at first order in $m_\phi^2$.

\subsection{Cancellation in the mass basis}
\label{sec:mass}

Thus far, we have seen the extent to which symmetries control the form of possible contributions to the dipole moments of a massless Dirac fermion in a toy model where this fermion is accompanied by heavy neutral fermions and a massless charged scalar. We now map the germane parts of the full model in \cref{eq:model} onto this toy model. Naively, this will only result in a vanishing dipole moment if the full model orchestrates additional correlations between the parameters of the toy model appearing in \cref{eq:delta_d}. As we will see, this is precisely what occurs.

Setting the Higgs to its vev and rotating to the mass basis in the fermion sector, the interactions in \cref{eq:model} that involve $e$ and $e^c$, the charged Higgs, as well as the neutral components of the massive fermions are those of \cref{eq:toy}, after making the identification $\phi \rightarrow H^+$, $\{ \psi, \psi^c \} \rightarrow \{ L^0, {L^c}^0 \}$ and $\{ \hat \psi, \hat \psi^c \} \rightarrow \{ S, S^c\}$. In terms of the parameters in \cref{eq:model}, the Yukawa couplings and mass terms appearing in \cref{eq:toy} are given by
\begin{align} \label{eq:fulltotoy1}
& m = m_L^{} , & & y_L^{} = Y_L^{} \,{\theta^c}^* , & & y_R^* = - Y_R^{} , \\ \label{eq:fulltotoy2}
& \hat m = m_S^{} , & & {\hat y}_L = Y_L^{} , & & {\hat y}_R^* = Y_R^{} \,\theta^* .
\end{align}
The $\theta$ and $\theta^c$ factors are mixing angles among the various fermions. Explicitly, they are defined by the flavor-to-mass eigenbasis rotation as follows:
\begin{equation}
\begin{pmatrix} L^0 \\ S \end{pmatrix} \rightarrow \begin{pmatrix} 1 & -\theta^* \\ \theta & 1 \end{pmatrix} \begin{pmatrix} L^0 \\ S \end{pmatrix} \qquad \text{and} \qquad
\begin{pmatrix} {L^c}^0 \\ S^c \end{pmatrix} \rightarrow \begin{pmatrix} 1 & -\theta^c \\ {\theta^c}^* & 1 \end{pmatrix} \begin{pmatrix} {L^c}^0 \\ S^c \end{pmatrix} .
\end{equation}
To $\mathcal{O} ( v / m_L)$ and $\mathcal{O} (v / m_S)$ they are given by
\begin{align} \label{eq:mixingangles}
\theta = \frac{v}{\sqrt{2}} \frac{m_S^* Y_V'}{|m_L^{}|^2 - |m_S^{}|^2} \qquad \qquad \text{and} \qquad \qquad \theta^c = \frac{v}{\sqrt{2}} \frac{m_L^* Y_V'}{|m_L^{}|^2 - |m_S^{}|^2}.	
\end{align}
Up to overall multiplicative factors, the contribution to the electromagnetic dipole moment from each individual massive fermion satisfies
\begin{equation}
e \,\frac{y_L^{} y_R^*}{m} = - e\, \frac{{\hat y}_L^{} {\hat y}_R^*}{\hat m} = - \frac{v}{\sqrt{2}} \frac{e \, Y_L^{} Y_R^{} Y'^*_V}{|m_L^{}|^2 - |m_S^{}|^2} \,,
\end{equation}
and the combination in \cref{eq:delta_d} therefore vanishes.

In the mass eigenbasis, the absence of what our spurion analysis identifies as the leading contribution to the dipole term comes down to the mixing angles in the fermion sector satisfying the following relationship:
\begin{equation}
\frac{\theta^c}{\theta} = \frac{m_L^*}{m_S^*} .
\label{eq:magic}
\end{equation}
In each of the diagrams of \cref{fig:oneloop_dipole}, the mass scaling of the mixing angles cancels against the fermion mass factor appearing in the propagator, making it possible for the two diagrams to cancel --- despite the two fermions propagating inside the loop appearing at different scales.

Is there a symmetry enforcing this relationship? Before rotating to the mass eigenbasis, the mass terms in the Lagrangian involving the neutral fermions read
\begin{align}
\mathcal{L} \supset	& \left\{ - m_L^{} L^0 {L^c}^0 - m_S^{} S S^c - \frac{Y'_V v}{\sqrt{2}} L^0 S^c \right\} + \text{h.c.} 
\label{eq:L_flavor}
\end{align}
To identify how the mixing angles are related notice that \cref{eq:L_flavor}, together with the corresponding kinetic terms, is invariant under a flavor symmetry $C'$ acting as \footnote{The notation $C^{\prime}$ shows a similarity to the charge conjugation $C$ which also interchanges two Weyl fields.}
\begin{equation}
C' \, : \qquad L^0 \leftrightarrow S^c \qquad \text{and} \qquad {L^c}^0 \leftrightarrow S ,
\label{eq:GC_def}
\end{equation}
provided it also acts (spuriously) on the mass parameters as follows:
\begin{equation}
C' \, : \qquad m_L^{} \leftrightarrow m_S^{} .
\label{eq:GC_spurions}
\end{equation}
In turn, this implies that the two mixing angles must satisfy the following relationship:
\begin{equation}
\theta^c = - \theta \Big|_{m_S^{} \leftrightarrow m_L^{}} .
\label{eq:GC_epsilons}
\end{equation}

Let us now turn to the form of these mixing angles, focusing on $\theta$ first. As a mixing angle, it must be proportional to $1 / (|m_L|^2 - |m_S|^2)$. The combination of couplings/masses carrying the appropriate $U(1)$ charge assignments is given by
\begin{equation}
\theta = N\, \frac{m_S^* Y'_V v / \sqrt{2}}{|m_L^{}|^2 - |m_S^{}|^2} ,
\label{eq:eps}
\end{equation}
where $N$ is an overall coefficient that we cannot obtain with a spurion analysis. Our flavor symmetry therefore tells us that the other mixing angle must be:
\begin{equation}
\theta^c = N\, \frac{m_L^* Y'_V v / \sqrt{2}}{|m_L^{}|^2 - |m_S^{}|^2} .
\label{eq:epsc}
\end{equation}
\Cref{eq:eps,eq:epsc} clearly satisfy \cref{eq:magic}, leading to a vanishing dipole.

\subsection{Surprise or expectation?}

Let us go back to the flavor basis, before setting the Higgs to its vev. The relevant interactions read:
\begin{align}
	\mathcal{L} \supset	& \left\{ - m_L^{} L^0 {L^c}^0 - m_S S S^c - Y'_V H^0 L^0 S^c + Y_L^{} H^+ e S^c - Y_R^{} H^- L^0 e^c \right\} + \text{h.c.} 
\label{eq:L_flavor_more}
\end{align}
These are not all the interactions appearing in the model of \cref{eq:model} --- just those that participate in the one-loop diagram of \cref{fig:oneloop_2}(b) (or, equivalently, those in \cref{fig:oneloop_dipole}).

The flavor symmetry introduced in the previous section can be extended to include the SM field content into what we will refer to as a generalized version of charge conjugation symmetry. By definition, under $C'$ all the SM fields, 
except the neutrino which does not enter in our analysis, transform as they do under ordinary charge conjugation:
\begin{equation}
	C' \, : \qquad e \leftrightarrow e^c , \qquad  
	H^+ \leftrightarrow H^-, \qquad A_\mu \leftrightarrow - A_\mu ,
\end{equation}
and the list of spurious transformations in \cref{eq:GC_spurions} is now supplemented with:
\begin{equation}
	C' \, : \qquad Y_L \leftrightarrow - Y_R\,.
\end{equation}

So if we compute the dipole contribution from, say, the diagram of \cref{fig:oneloop_dipole}(a), should we be surprised that the result is entirely cancelled by the diagram of \cref{fig:oneloop_dipole}(b)? The contribution to the electromagnetic dipole moments from \cref{fig:oneloop_dipole}(a), corresponding to the fermion with mass $m_L^{}$ propagating inside the loop, is given by
\begin{equation}
	\tau_{(L)}^{} = \frac{e}{32 \pi^2} \frac{v}{\sqrt{2}} \frac{ Y_L^{} Y_R^{} Y'^*_V }{|m_L^{}|^2 - |m_S^{}|^2} \,.
\label{eq:tauL}
\end{equation}
Overall, $\tau$ as the coefficient of the dipole operator in the effective action (second term of \cref{Eq:massanddipole1}) must be invariant under the action of $C'$.
However,
\begin{equation}
	\tau_{(L)}^{} \xrightarrow{C'} \frac{e}{32 \pi^2} \frac{v}{\sqrt{2}} \frac{ Y_L^{} Y_R^{} Y'^*_V }{|m_S^{}|^2 - |m_L^{}|^2} = - \tau_{(L)} \ .
\end{equation}
Since the contribution to $\tau$ from the diagram of \cref{fig:oneloop_dipole}(a) is odd under $C'$, it must be entirely cancelled by the diagram in \cref{fig:oneloop_dipole}(b).
The cancellation between the two diagrams in \cref{fig:oneloop_dipole}, featuring fermions with different masses propagating inside the loop, is therefore not so much a surprise but an expectation.

Alternatively, one could have defined the action of a generalized parity symmetry $P'$ that exchanges $L^0 \leftrightarrow {S^c}^\dagger$ and ${L^c}^0 \leftrightarrow S^\dagger$ (in combination with spatial reflection), and that acts as usual on the field content of the SM fields. As spurions, $P'$ exchanges $m_L^{} \leftrightarrow m_S^*$, $Y'_V \leftrightarrow Y'^*_V$ and $Y_L^{} \leftrightarrow Y_R^*$. In turn, $\mathbb{Re}(\tau_L^{}) \rightarrow - \mathbb{Re}(\tau_L^{})$ whereas $\mathbb{Im}(\tau_L^{}) \rightarrow \mathbb{Im}(\tau_L^{})$ under $P'$. Since magnetic and electric dipole moments are respectively even and odd under parity symmetry, one immediately concludes that the diagram of \cref{fig:oneloop_dipole}(b) must entirely cancel the contribution to both magnetic and electric dipole moments coming from the diagram of \cref{fig:oneloop_dipole}(a).

By contrast, the contribution to the SM fermion mass from one of the two massive fermions --- corresponding to one of the two terms inside the parenthesis in \cref{eq:deltammu} --- is neither even nor odd under $C'$. The contribution from the additional fermion cancels the odd part, leaving $\delta m (\mu) \xrightarrow{C'} \delta m (\mu)$, as can be checked explicitly from \cref{eq:deltammu}.

\section{Symmetries at work: gauge basis}
\label{sec:unbroken}

We can gain further insight into the role of symmetries by performing a spurion analysis in the gauge basis, {\it i.e.}\ in the unbroken phase of electroweak symmetry. 
The mass basis calculation in the previous section hints at a $\mathbb{Z}_2$ symmetry relating fields that belong to $SU(2)_L^{}$ doublets and those that are $SU(2)_L^{}$ singlets, {\it e.g.}\ $L^0$ and $S^c$, as the key ingredient underlying a symmetry explanation for the vanishing dipole moment. 
In the original model, \cref{eq:model}, manifesting this symmetry then inevitably breaks $SU(2)_L^{}$. In this section, we take an alternative perspective: by embedding \cref{eq:model} in an $SU(2)_L^{}\times SU(2)_R^{}$ symmetric theory, we will see that the dipole moment inherits symmetry protection from the larger theory. This requires introducing a discrete symmetry, $P_{LR}^{}$, that relates the two $SU(2)$'s and plays a similar role to $C'$ in \cref{sec:symmetry}.
The freedom of performing field redefinitions will prove powerful in this analysis.

\subsection{$SU(2)_L^{}\times SU(2)_R^{}$ embedding}

We begin by introducing auxiliary Weyl fermion fields, $\nu^c$, $E^c$ and $E$, which combine with the $SU(2)_L^{}$ singlets in the model of \cref{eq:model} to form $SU(2)_R^{}$ doublets. 
The field content of this extended theory therefore includes $SU(2)_L^{}$ doublet and $SU(2)_R^{}$ doublet fermions that are pairwise related by $P_{LR}^{}$:
\begin{equation}
\begin{array}{ccccccc}
&
l^i &= 
\begin{pmatrix}
\nu \\ e
\end{pmatrix} ,\qquad
& L^i & = 
\begin{pmatrix}
L^0 \\ L^-
\end{pmatrix} ,\qquad
& 
L^c_i & = 
\begin{pmatrix}
-L^{c0} \\ L^{c+}
\end{pmatrix} 
,\\
P_{LR}^{}\quad & \updownarrow & & \updownarrow & & \updownarrow &
\\
&
r^c_{i'} &= 
\begin{pmatrix}
\nu^c \\ e^c
\end{pmatrix} ,\qquad
& R^c_{i'} & = 
\begin{pmatrix}
S^c \\ E^c
\end{pmatrix} ,\qquad
& 
R^{i'} & =
\begin{pmatrix}
-S \\ E
\end{pmatrix} 
.
\end{array}
\end{equation}
The Higgs scalar is embedded in the bifundamental in the familiar way:
\begin{equation}
\Sigma\indices{^i_{j'}} = \bigl( \Sigma\indices{^i_1} ,\; \Sigma\indices{^i_2}\bigr) 
= \bigl( \widetilde H^i ,\; H^i \bigr) 
= \begin{pmatrix}
H^{0*} & H^+ \\
-H^- & H^0
\end{pmatrix} ,
\end{equation}
where $\widetilde{H}^i \equiv \epsilon^{ij} H^*_j$, $H^- \equiv (H^+)^*$. 
We define $P_{LR}^{}$ to act on the Higgs field as follows:
\begin{equation}
P_{LR}^{}:\quad \Sigma\indices{^i_{j'}} \leftrightarrow \Sigma\indices{^j_{i'}} \quad (\text{{\it i.e.}} \;\; H^+ \leftrightarrow -H^-).
\end{equation}
Throughout this section, we use unprimed and primed lowercase letters for $SU(2)_L^{}$ and $SU(2)_R^{}$ (anti-)fundamental indices, respectively, and will use capital letters to refer to them collectively ({\it e.g.}\ $I = i, i'$).
Our choice of upper vs.\ lower indices in the equations above is purely a matter of convention (with a minus sign introduced in $L^c_i$ to be consistent with conventions in the previous sections where $L^{c\,i} = (L^{c+},\, L^{c0})^T$); as usual, these indices can be raised/lowered with $\epsilon^{IJ}$, $\epsilon_{IJ}^{}$, where $\epsilon^{12} = -\epsilon^{21} = -\epsilon_{12}^{} = \epsilon_{21}^{} = 1$.  
Note that $P_{LR}^{}$ takes an $SU(2)_L^{}$ fundamental (unprimed upper index) to an $SU(2)_R^{}$ anti-fundamental (primed lower index), etc. 
Complex conjugation (c.c.)\ also raises/lowers indices, as it takes a fundamental to an anti-fundamental, and vice versa; to fix the sign convention consistent with $(\epsilon_{IJ}^{})^* = -\epsilon^{IJ}$, we attach a plus (minus) sign when c.c.\ lowers (raises) an index, {\it e.g.}\ $H^i \overset{\text{c.c.}}{\longrightarrow}H^*_i$, whereas $H_i = \epsilon_{ij}^{} H^j \overset{\text{c.c.}}{\longrightarrow} -\epsilon^{ij} H^*_j = -H^{*i}$. 
Note that $\Sigma\indices{^i_{j'}}$ and its conjugate are not independent: 
\begin{equation}
{\Sigma^*}\indices{^i_{j'}} = \epsilon^{ik} \epsilon_{j'l'}^{} {\Sigma^*}\indices{_k^{\,l'}}
= -\epsilon^{ik} \epsilon_{j'l'}^{} \bigl(\Sigma\indices{^k_{l'}}\bigr)^* = 
\begin{pmatrix}
H^{0*} & H^+ \\
-H^- & H^0
\end{pmatrix}
= \Sigma\indices{^i_{j'}} \,.
\label{eq:Sigma_cc}
\end{equation}

We can write the Lagrangian of this extended theory in a compact form by introducing the following field multiplets that are interchanged under $P_{LR}^{}$:
\begin{equation}
\Psi^I \equiv 
\begin{pmatrix}
l^i \\ L^i \\ R^{i'}
\end{pmatrix} 
\qquad \xleftrightarrow{\;P_{LR}^{}\;}\qquad
\Psi^c_I \equiv
\begin{pmatrix}
r^c_{i'} \\ R^c_{i'} \\ L^c_i
\end{pmatrix} \,.
\end{equation}
It follows that
\begin{equation}
\mathcal{L} \supset 
i\, \Psi^\dagger_I \bar\sigma^\mu D_\mu \Psi^I
+i\, \Psi^{c\dagger I} \bar\sigma^\mu D_\mu \Psi^c_I 
- \Psi^{\T I} \bigl( \Mat{M} + \Mat{U}\bigr)\indices{_I^J}\, \Psi^c_J  \,,
\label{eq:model_embed}
\end{equation}
where
\begin{align}
&
\Mat{M} = 
\begin{pmatrix}
0 & 0 & 0 \\
0 & 0 & \mat{m}_L^{} \\
0 & \mat{m}_R^{} & 0
\end{pmatrix} \,,\qquad
\Mat{U} = 
\begin{pmatrix}
0 & \mat{U}_{lR}^{} & 0 \\
\mat{U}_{Lr}^{} & \mat{U}_{LR}^{} & 0 \\
0 & 0 & \mat{U}_{RL}^{}
\end{pmatrix} \,,\\[6pt]
&
\bigl[\mat{U}_{lR}^{} \bigr]\indices{_i^{\,j'}} = \bigl[\mat{y}_L^{}\bigr]\indices{_i^{\,j'}_k^{\,l'}} \,\Sigma\indices{^k_{l'}} \,,\qquad
\bigl[\mat{U}_{Lr}^{} \bigr]\indices{_i^{\,j'}} = \bigl[\mat{y}_R^{}\bigr]\indices{_i^{\,j'}_k^{\,l'}} \,\Sigma\indices{^k_{l'}} \,,\label{eq:U_1}\\
&
\bigl[\mat{U}_{LR}^{} \bigr]\indices{_i^{\,j'}} = \bigl[\mat{y}_V'\bigr]\indices{_i^{\,j'}_k^{\,l'}} \,\Sigma\indices{^k_{l'}} \,,\qquad
\bigl[\mat{U}_{RL}^{} \bigr]\indices{_{i'}^j} = \bigl[\mat{y}_V^{}\bigr]\indices{_{i'}^j_{\,k}^{\,l'}} \,\Sigma\indices{^k_{l'}} \,.\label{eq:U_2}
\end{align}
The mass and Yukawa tensors have the following nonzero components:
\begin{align}
&
\bigl[ \mat{m}_L^{} \bigr]\indices{_i^{\,j}} = \delta_i^j\, m_L^{} \,,\qquad
\bigl[ \mat{m}_R^{} \bigr]\indices{_1^1} = m_S^{} \,,\quad
\bigl[ \mat{m}_R^{} \bigr]\indices{_2^2} = m_E^{}\; (\to\infty) \,,\label{eq:phys_m}\\[2pt]
&
\bigl[ \mat{y}_L^{} \bigr]\indices{_i^1_k^2} = -\epsilon_{ik}^{}\, Y_L^{} \,,\quad
\bigl[ \mat{y}_R^{} \bigr]\indices{_i^2_k^1} = \epsilon_{ik}^{}\, Y_R^{} \,,\quad
\bigl[ \mat{y}_V' \bigr]\indices{_i^1_k^2} = -\epsilon_{ik}^{}\, Y_V' \,,\quad
\bigl[ \mat{y}_V^{} \bigr]\indices{_1^j_k^1} = \delta_k^j\, Y_V^{} \,.
\label{eq:phys_y}
\end{align}
We have flipped the sign conventions of $m_L^{}$, $m_S^{}$ compared to the previous sections for convenience.
Note that the auxiliary fermions $\nu^c$, $E^c$, $E$ do not participate in the Yukawa interactions, so the extended model \cref{eq:model_embed} is equivalent to the original one in \cref{eq:model} as far as observables involving physical (non-auxiliary) fields are concerned.

The Lagrangian \cref{eq:model_embed} is invariant under $SU(2)_L^{}\times SU(2)_R^{}$ when the mass and Yukawa tensors are treated as spurions transforming according to their indices. 
It is also $P_{LR}^{}$ invariant if we define the latter to act spuriously as $\Mat{M}\indices{_I^J} \leftrightarrow {\Mat{M}^\T}\indices{_J^{\,I}}$, $\Mat{U}\indices{_I^J} \leftrightarrow {\Mat{U}^\T}\indices{_J^{\,I}}$; in other words,
\begin{align}
P_{LR}:\qquad &
\bigl[ \mat{m}_L^{}\bigr]\indices{_i^{\,j}} \leftrightarrow \bigl[ \mat{m}_R^{} \bigr]\indices{_{j'}^{i'}} \,,\nonumber\\
&
\bigl[ \mat{y}_L^{} \bigr]\indices{_i^{\,j'}_k^{\,l'}} \leftrightarrow \bigl[ \mat{y}_R^{} \bigr]\indices{_j^{\,i'}_l^{\,k'}} \,,\quad
\bigl[ \mat{y}_V' \bigr]\indices{_i^{\,j'}_k^{\,l'}} \leftrightarrow \bigl[ \mat{y}_V' \bigr]\indices{_j^{\,i'}_l^{\,k'}} \,,\quad
\bigl[ \mat{y}_V^{} \bigr]\indices{_{i'}^j_{\,k}^{\,l'}} \leftrightarrow \bigl[ \mat{y}_V^{} \bigr]\indices{_{j'}^i_{\,l}^{\,k'}}\,.
\label{eq:PLR_m_y}
\end{align}

We also need to introduce a spurion for the $U(1)_\text{EM}^{}$ coupling of the Higgs field. 
The relevant part of covariant derivative reads $D_\mu\Sigma\indices{^i_{j'}} = \partial_\mu \Sigma\indices{^i_{j'}} -ieA_\mu \,Q\indices{^i_{j'k}^{\,l'}}\, \Sigma\indices{^k_{l'}}$, where $Q\indices{^i_{j'k}^{\,l'}}$ is the $U(1)_\text{EM}^{}$ generator with the following nonzero components:
\begin{equation}
Q\indices{^1_{21}^2} = -Q\indices{^2_{12}^1} = 1\,;\qquad
\text{equivalently,}\quad
Q\indices{_2^1_1^2} = -Q\indices{_1^2_2^1} =
Q\indices{^1_2^2_1} = -Q\indices{^2_1^1_2} = 1 \,.
\end{equation}
Hermiticity of $Q$ implies $\bigl(Q\indices{^i_{j'k}^{\,l'}}\bigr)^* = Q\indices{^k_{l'i}^{\,j'}}$, and it follows that
\begin{equation}
\bigl|D_\mu\Sigma\indices{^i_{j'}}\bigr|^2 \supset ieA_\mu \bigl(Q\indices{_i^{\,j'}_k^{\,l'}} - Q\indices{_k^{\,l'}_i^{\,j'}}\bigr) \,\bigl(\partial^\mu \Sigma\indices{^i_{j'}}\bigr)\, \Sigma\indices{^k_{l'}}
= 2ieA_\mu\,Q\indices{_i^{\,j'}_k^{\,l'}} \,\bigl(\partial^\mu \Sigma\indices{^i_{j'}}\bigr)\, \Sigma\indices{^k_{l'}} \,,
\label{eq:Sigma_EM}
\end{equation}
where we have used \cref{eq:Sigma_cc} and the antisymmetry of $Q\indices{_i^{\,j'}_k^{\,l'}}$. 
For \cref{eq:Sigma_EM} to be invariant under $P_{LR}^{}$, we define the latter to act on the $U(1)_\text{EM}^{}$ charge spurion as:
\begin{equation}
P_{LR}^{}: \qquad Q\indices{_i^{\,j'}_k^{\,l'}} \leftrightarrow Q\indices{_j^{\,i'}_l^{\,k'}} \,.
\label{eq:PLR_Q}
\end{equation}
We can similarly introduce $U(1)_\text{EM}^{}$ charge spurions for the fermion fields, which however will not be needed in our calculation of dipole moments because, as we will see in \cref{sec:explain}, charged fermions do not enter the relevant one-loop diagrams.

\subsection{Field redefinition}
\label{sec:redef}

With the $SU(2)_L^{}\times SU(2)_R^{} \times P_{LR}^{}$ symmetric theory written down in the previous subsection, one might be tempted to immediately enumerate products of spurions that have the right transformation properties to contribute to dipole moments. 
However, this is complicated by the fact that the corresponding loop diagrams involve nondegenerate masses from both $\mat{m}_L^{}$ and $\mat{m}_R^{}$, resulting in nontrivial mass dependence of the loop integral. 
Obviously, this is due to the presence of Yukawa couplings (in the $\Mat{U}$ matrix) connecting nondegenerate fermions. 
The trick we are going to play next is to remove these couplings via a field redefinition.

Consider a field redefinition of the following form:
\begin{equation}
\Psi^I \to {\Mat{O}^*}\indices{^I_J} \,\Psi^J \,,\qquad
\Psi^c_I \to {\Mat{O}^c}\indices{_I^J} \,\Psi^c_J\,,
\end{equation}
where $\Mat{O}$, $\Mat{O}^c$ are Higgs field-dependent $3\times 3$ unitary matrices in field multiplet space. 
We choose $\Mat{O}$, $\Mat{O}^c$ such that
\begin{equation}
\bigl[\Mat{O}^\dagger (\Mat{M} + \Mat{U}) \,\Mat{O}^c\bigr]\indices{_I^L} =
{\Mat{O}^\dagger}\indices{_I^J} (\Mat{M} + \Mat{U})\indices{_J^K} \, {\Mat{O}^c}\indices{_K^L} = 
\begin{pmatrix}
\;\;\cdot\;\; & 0 & 0 \\
0 & 0 & \;\;\cdot\;\; \\
0 & \;\;\cdot\;\; & 0
\end{pmatrix} \,,
\label{eq:redef_diag}
\end{equation}
where ``$\,\cdot\,$'' denotes a (Higgs field-dependent) nonzero entry whose detailed form will not be needed. 
This field redefinition takes us to an operator basis where zero-derivative operators ({\it e.g.}\ Yukawa couplings) involving nondegenerate fermions are absent --- it is essentially the unbroken phase version of ``going to the mass basis;'' when the Higgs is set to its vev, it reproduces the diagonalization of mass matrix in the broken phase. 
We can solve \cref{eq:redef_diag} order by order in $\Mat{U}\sim\mathcal{O}(\mat{y}\Sigma)$, writing 
\begin{equation}
\Mat{O} = \Mat{1} +\Mat{O}_{(1)} + \Mat{O}_{(2)} +\dots  \,,\qquad
\Mat{O}^c = \Mat{1} +\Mat{O}^c_{(1)} + \Mat{O}^c_{(2)} +\dots  \,,
\end{equation}
with $\Mat{O}_{(n)},\, \Mat{O}^c_{(n)} \sim \mathcal{O}\bigl((\mat{y}\Sigma)^n\bigr)$. 
We present the solution for $n=1,2$ in \cref{sec:app_redef}.

Meanwhile, from the gauged kinetic terms in the original basis we obtain the following one-derivative operators upon field redefinition:
\begin{align}
i\, \Psi^\dagger_I \bar\sigma^\mu D_\mu \Psi^I &\to 
i\, \Psi^\dagger_I \bar\sigma^\mu D_\mu \Psi^I
+i\, \Psi^\dagger_I \bar\sigma^\mu {\bigl[ \Mat{O}^\dagger (D_\mu\Mat{O}) \bigr]^*}\indices{^I_{J}}\, \Psi^J \,,\label{eq:kin_redef_1}\\[4pt]
i\, \Psi^{c\dagger I} \bar\sigma^\mu D_\mu \Psi^c_I &\to
i\, \Psi^{c\dagger I} \bar\sigma^\mu D_\mu \Psi^c_I
+i\, \Psi^{c\dagger I} \bar\sigma^\mu \bigl[ \Mat{O}^{c\dagger}(D_\mu\Mat{O}^c)\bigr]\indices{_I^{\,J}} \,\Psi^c_J\,.\label{eq:kin_redef_2}
\end{align}
Substituting in $\Mat{O}$, $\Mat{O}^c$ up to the second order (see \cref{sec:app_redef}), we find the following dimension-five and dimension-six operators in the new basis:
\begin{align}
\mathcal{L}_\text{dim-5} \supset &\;
-i\, \bigl[ \mat{c}_{5L}^{}\bigr]\indices{_i^{\,j'}_k^{\,l'}} \bigl(D_\mu \Sigma\indices{^k_{l'}}\bigr)\, R^\dagger_{j'} \,\bar\sigma^\mu\, l^i
-i\,\bigl[ \mat{c}_{5R}^{}\bigr]\indices{_i^{\,j'}_k^{\,l'}} \bigl(D_\mu \Sigma\indices{^k_{l'}} \bigr)\, L^{c\dagger i}\,\bar\sigma^\mu \,r^c_{j'} +\text{h.c.} \,, \label{eq:L_dim5}\\
\mathcal{L}_\text{dim-6} \supset &\;
i\, \bigl[ \mat{c}_{6L}^{}\bigr]\indices{_i^{\,j}_k^{\,l'}_m^{\,n'}} \,\Sigma\indices{^k_{l'}} \,\bigl(D_\mu \Sigma\indices{^m_{n'}}\bigr) \, L^\dagger_j \,\bar\sigma^\mu \,l^i
\nonumber\\
&+i\, \bigl[ \mat{c}_{6R}^{}\bigr]\indices{_{i'}^{j'}_k^{\,l'}_m^{\,n'}} \,\Sigma\indices{^k_{l'}} \,\bigl(D_\mu \Sigma\indices{^m_{n'}}\bigr) \, R^{c\dagger i'} \,\bar\sigma^\mu \,r^c_{j'} +\text{h.c.}\,.\label{eq:L_dim6}
\end{align}
The operator coefficients are given by
\begin{align}
\bigl[ \mat{c}_{5L}^{} \bigr]\indices{_i^{\,j'}_k^{\,l'}} =&\; \bigl[ \mat{y}_L^{} \mat{m}_R^{-1} \bigr]\indices{_i^{\,j'}_k^{\,l'}} \,,\qquad
\bigl[ \mat{c}_{5R}^{} \bigr]\indices{_i^{\,j'}_k^{\,l'}} = \bigl[ \mat{m}_L^{-1} \,\mat{y}_R^{} \bigr]\indices{_i^{\,j'}_k^{\,l'}} \,,
\label{eq:c5}\\[6pt]
\bigl[ \mat{c}_{6L}^{} \bigr]\indices{_i^{\,j}_k^{\,l'}_m^{\,n'}} =&\; \bigl[ \mat{y}_L^{} \bigr]\indices{_i^{\,p'}_k^{\,l'}} \bigl[ \mat{m}_R^{-1}\, \mat{y}_V^{} \mat{m}_L^{-1} \bigr]\indices{_{p'}^j_{\,m}^{\,n'}} \nonumber\\[2pt]
&\; +\bigl[ \mat{y}_L^{} \bigr]\indices{_i^{\,p'}_m^{\,n'}} \bigl[ \mat{m}_R^{-1} \bigl(\mat{\Delta \bar m}^{-2}\bigr) \mat{m}_R^{} \bigr]\indices{_r^{\,j}_{p'}^{q'}} \bigl[ \mat{y}_V^{\prime\dagger} +\mat{m}_R^\dagger \mat{y}_V^{} \mat{m}_L^{-1} \bigr]\indices{_{q'}^r_k^{\,l'}} \,,
\label{eq:c6L}\\[6pt]
\bigl[ \mat{c}_{6R}^{} \bigr]\indices{_{i'}^{j'}_k^{\,l'}_m^{\,n'}} =&\; \bigl[ \mat{m}_R^{-1} \,\mat{y}_V^{} \mat{m}_L^{-1} \bigr]\indices{_{i'}^p_m^{\,n'}} \bigl[ \mat{y}_R^{} \bigr]\indices{_p^{\,j'}_k^{\,l'}} \nonumber\\
&\; -\bigl[ \mat{m}_L^{} \bigl(\mat{\Delta m}^{-2}\bigr) \mat{m}_L^{-1} \bigr]\indices{_q^{\,p}_{i'}^{r'}} \bigl[ \mat{y}_V^{\prime\dagger} +\mat{m}_R^{-1}\, \mat{y}_V^{} \mat{m}_L^\dagger \bigr]\indices{_{r'}^q_k^{\,l'}} \bigl[ \mat{y}_R^{} \bigr]\indices{_p^{\,j'}_m^{\,n'}} \,,
\label{eq:c6R}
\end{align}
where $\bigl[ \mat{y}_V^{\prime\dagger} \bigr]\indices{_{q'}^r_{k}^{\,l'}} \equiv \bigl[ \mat{y}_V^{\prime *} \bigr]\indices{^r_{q'k}^{\,l'}} = \Bigl(\bigl[ \mat{y}_V^{\prime} \bigr]\indices{_r^{\,q'k}_{l'}}\Bigr)^* = \Bigl(\epsilon^{ks}\,\epsilon_{l't'}\,\bigl[ \mat{y}_V^{\prime} \bigr]\indices{_r^{\,q'}_s^{\,t'}}\Bigr)^*$, and $\mat{\Delta m}^{-2}$, $\mat{\Delta \bar m}^{-2}$ are inverse tensors of
\begin{align}
\bigl[ \mat{\Delta m}^2 \bigr]\indices{_i^{\,j}_{\,k'}^{l'}} &\equiv \delta_i^j \,\bigl[\mat{m}_R^\dagger\mat{m}_R^{}\bigr]\indices{_{k'}^{l'}} -\bigl[\mat{m}_L^\dagger \mat{m}_L^{} \bigr]\indices{_i^{\,j}} \,\delta_{k'}^{l'} \,,\label{eq:dm2}\\
\bigl[ \mat{\Delta \bar m}^2 \bigr]\indices{_i^{\,j}_{\,k'}^{l'}} &\equiv \delta_i^j \,\bigl[\mat{m}_R^{}\mat{m}_R^\dagger\bigr]\indices{_{k'}^{l'}} -\bigl[\mat{m}_L^{} \mat{m}_L^\dagger \bigr]\indices{_i^{\,j}} \,\delta_{k'}^{l'} \,,\label{eq:dmbar2}
\end{align}
respectively. 
Contraction between $SU(2)_{L,R}^{}$ indices carried by the fermions has been left implicit where it can be unambiguously recovered from matrix multiplication, {\it e.g.}\ $\bigl[ \mat{y}_L^{} \mat{m}_R^{-1} \bigr]\indices{_i^{\,j'}_k^{\,l'}}  = \bigl[ \mat{y}_L^{} \bigr]\indices{_i^{\,m'}_k^{\,l'}} \bigl[ \mat{m}_R^{-1} \bigr]\indices{_{m'}^{j'}}$, $\bigl[ \mat{m}_R^{-1}\, \mat{y}_V^{} \mat{m}_L^{-1} \bigr]\indices{_{p'}^j_{\,m'}^{n'}} = \bigl[ \mat{m}_R^{-1} \bigr]\indices{_{p'}^{q'}} \bigl[ \mat{y}_V^{} \bigr]\indices{_{q'}^r_{\,m'}^{n'}} \bigl[ \mat{m}_L^{-1} \bigr]\indices{_r^{\,j}}$. 
We only presented operators involving both heavy and light fermions above; additional operators that come from expanding \cref{eq:kin_redef_1,eq:kin_redef_2} ({\it e.g.}\ those involving only heavy fermions and the Higgs) do not contribute to the dipole at one-loop level.

The field redefinition we have performed in this subsection preserves $P_{LR}^{}$, under which ${\Mat{O}^*}\indices{^I_J} \leftrightarrow {\Mat{O}^c}\indices{_I^{\,J}}$. 
Using \cref{eq:PLR_m_y} and noting that $P_{LR}^{}$ exchanges $\bigl[ \mat{\Delta m}^{-2} \bigr]\indices{_i^{\,j}_{\,k'}^{l'}} \leftrightarrow -\bigl[\mat{\Delta \bar m}^{-2}\bigr]\indices{_l^{\,k}_{\,j'}^{i'}}$, we can readily check that 
\begin{equation}
P_{LR}^{}:\qquad 
\bigl[ \mat{c}_{5L} \bigr]\indices{_i^{\,j'}_k^{\,l'}} \leftrightarrow \bigl[ \mat{c}_{5R} \bigr]\indices{_j^{\,i'}_l^{\,k'}} \,,\qquad
\bigl[ \mat{c}_{6L} \bigr]\indices{_i^{\,j}_k^{\,l'}_m^{\,n'}} \leftrightarrow \bigl[ \mat{c}_{6R} \bigr]\indices{_{j'}^{i'}_l^{\,k'}_n^{\,m'}} \,.
\end{equation}
These are exactly the desired transformations for \cref{eq:L_dim5,eq:L_dim6} to be $P_{LR}^{}$-invariant.

\subsection{Explaining the magic zero}
\label{sec:explain}

In the new basis after the field redefinition discussed above, we have the following spurions:
\begin{equation}
\bigl[\mat{m}_L{}\bigr]\indices{_{\Li{i}}^{\Ri{\,j}}} \,,\;\; 
\bigl[\mat{m}_R^{}\bigr]\indices{_{\Ri{i'}}^{\Li{j'}}} \,,\;\;
Q\Hi{\indices{_i^{\,j'}_k^{\,l'}}} \,,\;\; 
\bigl[\mat{c}_{5L}^{}\bigr]\indices{_{\li{i}}^{\,\Ri{j'}}_{\Hi{k}}^{\,\Hi{l'}}} \,,\;\;
\bigl[\mat{c}_{5R}^{}\bigr]\indices{_{\Ri{i}}^{\,\li{j'}}_{\Hi{k}}^{\,\Hi{l'}}} \,,\;\;
\bigl[\mat{c}_{6L}^{}\bigr]\indices{_{\li{i}}^{\,\Li{j}}_{\Hi{k}}^{\,\Hi{l'}}_{\Hi{m}}^{\,\Hi{n'}}} \,,\;\; 
\bigl[\mat{c}_{6R}^{}\bigr]\indices{_{\Li{i'}}^{\li{j'}}_{\Hi{k}}^{\,\Hi{l'}}_{\Hi{m}}^{\,\Hi{n'}}} \,.
\label{eq:spurions}
\end{equation}
To make it transparent that the Lagrangian is invariant under $SU(2)_L^{}\times SU(2)_R^{}$ transformations of each field ($\li{l}$,  $\li{r^c}$, $\Li{L}$, $\Li{R^c}$, $\Ri{L^c}$, $\Ri{R}$, $\Hi{\Sigma}$), we use different colors to keep track of individual $SU(2)_L^{}\times SU(2)_R^{}$ factors. 
In what follows, we use these spurions to construct the dimension-six operator in the effective action that corresponds to electromagnetic dipole moments, and show that it is constrained by $SU(2)_L^{}\times SU(2)_R^{}\times P_{LR}^{}$ to vanish at one-loop level. 
A similar spurion analysis can be performed for the dimension-four Yukawa operator; we show in \cref{sec:app_yukawa} that the correct (nonvanishing) form of the muon mass threshold correction is reproduced in this way. The difference in the fates of the dipole and Yukawa operators is due in part to their operator dimensions: the consequences of symmetry depend on the allowed powers of the heavy mass scales.

\begin{figure}[t]
	\centering
	\includegraphics[scale=1]{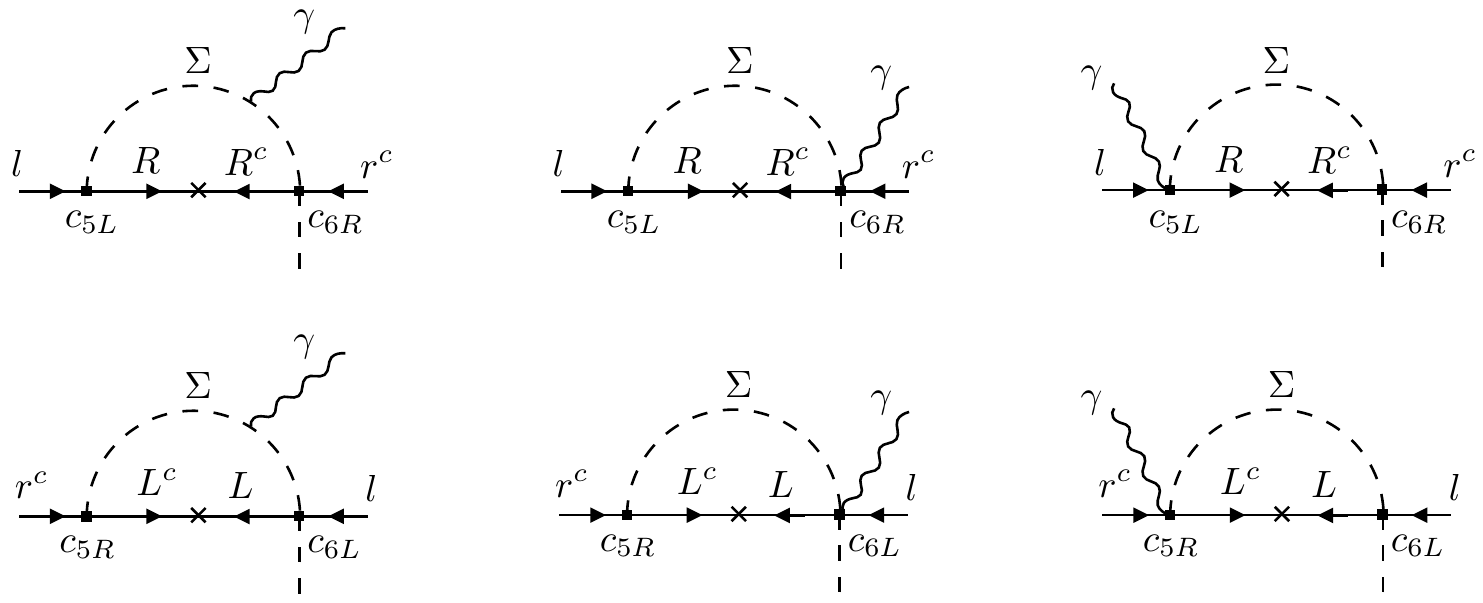}
	\caption{
		One-loop diagrams contributing to the the dipole operator in the $SU(2)_L^{}\times SU(2)_R^{}$ extended theory, in the basis after field redefinition. 
		The three diagrams in the first (second) row have the coupling structure shown in the first (second) term of \cref{eq:L_dipole} --- both terms are related by $P_{LR}^{}$ and are shown to cancel each other.
	}
	\label{fig:oneloop_newbasis}
\end{figure}

To construct the dipole, we need to contract products of the spurions in \cref{eq:spurions} with $e\,F^{\alpha\beta}\,l^{\li{i}}_\alpha r^c_{\beta\li{j'}}\,\Sigma\Hi{\indices{^k_{l'}}}$ (where $\alpha$, $\beta$ are spinor indices, see \cref{sec:U(2)}). 
The following requirements must be satisfied:
\begin{itemize}
	\item The operator coefficient has mass dimension $-2$. Also, at one-loop level, the dipole operator coefficient must involve three powers of Yukawa couplings.
	\item Only contractions between same-color indices that are both unprimed or both primed are allowed in order to preserve $SU(2)_L^{}\times SU(2)_R^{}$. Also, the $U(1)$ particle number associated with each fermion flavor must be preserved.
	\item The result must be even under $P_{LR}^{}$, which exchanges $SU(2)_L^{}$ (unprimed) and $SU(2)_R^{}$ (primed) indices while preserving their colors:
	\begin{align}
	P_{LR}^{}:\qquad&
	l^{\li{i}}_\alpha \leftrightarrow r^c_{\alpha\li{i'}} \,,\qquad
	\bigl[ \mat{m}_L^{}\bigr]\indices{_{\Li{i}}^{\,\Ri{j}}} \leftrightarrow \bigl[ \mat{m}_R^{} \bigr]\indices{_{\Ri{j'}}^{\Li{i'}}} \,,\qquad
	Q\Hi{\indices{_i^{\,j'}_k^{\,l'}}} \leftrightarrow Q\Hi{\indices{_j^{\,i'}_l^{\,k'}}} \,,\nonumber\\
	&
	\bigl[\mat{c}_{5L}^{}\bigr]\indices{_{\li{i}}^{\,\Ri{j'}}_{\Hi{k}}^{\,\Hi{l'}}} \leftrightarrow \bigl[\mat{c}_{5R}^{}\bigr]\indices{_{\Ri{j}}^{\,\li{i'}}_{\Hi{l}}^{\,\Hi{k'}}} \,,\qquad
	\bigl[\mat{c}_{6L}^{}\bigr]\indices{_{\li{i}}^{\,\Li{j}}_{\Hi{k}}^{\,\Hi{l'}}_{\Hi{m}}^{\,\Hi{n'}}} \leftrightarrow \bigl[\mat{c}_{6R}^{}\bigr]\indices{_{\Li{j'}}^{\li{i'}}_{\Hi{l}}^{\,\Hi{k'}}_{\Hi{n}}^{\,\Hi{m'}}} \,.
	\end{align}
	\item The external Higgs $\Sigma\Hi{\indices{^k_{l'}}}$ must be contracted with the first pair of Higgs indices carried by $\bigl[\mat{c}_{6L}^{}\bigr]\indices{_{\li{i}}^{\,\Li{j}}_{\Hi{k}}^{\,\Hi{l'}}_{\Hi{m}}^{\,\Hi{n'}}}$ or $\bigl[\mat{c}_{6R}^{}\bigr]\indices{_{\Li{i'}}^{\li{j'}}_{\Hi{k}}^{\,\Hi{l'}}_{\Hi{m}}^{\,\Hi{n'}}}$, because this is the only non-derivative interaction of the $U(1)_\text{EM}^{}$-neutral component of the Higgs field in the new basis that can contribute to the dipole.
\end{itemize}
Imposing all these constraints we find a unique possibility:\footnote{\,To be more explicit, the spinor indices work out as follows: $l_\alpha\,(\bar\sigma^\mu)^{\dot\gamma \alpha} \,\epsilon_{\dot\gamma\dot\delta}\,(\bar\sigma^\nu)^{\dot\delta\beta}\,r^c_\beta = l^\alpha\, (\sigma^\mu\bar\sigma^\nu)\indices{_\alpha^\beta} \,r^c_\beta = l^\alpha\, \bigl[ \eta^{\mu\nu}\,\delta_\alpha^\beta -2i\,(\sigma^{\mu\nu})\indices{_\alpha^\beta}\bigr] \,r^c_\beta$. The first term does not contribute to the dipole, while the second term will contract with $F_{\mu\nu}$ to yield $F\indices{_\alpha^\beta}$.}
\begin{align}
\mathcal{O}_\text{dipole} \propto &\;
e\,F^{\alpha\beta}\,l^{\li{i}}_\alpha\, \bigl[\mat{c}_{5L}^{}\bigr]\indices{_{\li{i}}^{\,\Ri{r'}}_{\Hi{p}}^{\,\Hi{q'}}} \bigl[ \mat{m}_R^{} \bigr]\indices{_{\Ri{r'}}^{\Li{s'}}} \bigl[\mat{c}_{6R}^{}\bigr]\indices{_{\Li{s'}}^{\li{j'}}_{\Hi{k}}^{\,\Hi{l'}}_{\Hi{m}}^{\,\Hi{n'}}}  r^c_{\beta\li{j'}}\,\Sigma\Hi{\indices{^k_{l'}}}\,Q\Hi{\indices{^m_{n'}^p_{q'}}} \nonumber\\
&+
e\,F^{\alpha\beta}\,r^c_{\alpha\li{i'}}\, \bigl[\mat{c}_{5R}^{}\bigr]\indices{_{\Ri{r}}^{\,\li{i'}}_{\Hi{q}}^{\,\Hi{p'}}} \bigl[ \mat{m}_L^{} \bigr]\indices{_{\Li{s}}^{\,\Ri{r}}} \bigl[\mat{c}_{6L}^{}\bigr]\indices{_{\li{j}}^{\Li{\,s}}_{\Hi{l}}^{\,\Hi{k'}}_{\Hi{n}}^{\,\Hi{m'}}}  l_\beta^{\li{j}}\,\Sigma\Hi{\indices{^l_{k'}}}\,Q\Hi{\indices{^n_{m'}^q_{p'}}} \,.
\label{eq:L_dipole}
\end{align}
The two terms are manifestly related by $P_{LR}^{}$, and correspond to diagrams shown in the two rows of \cref{fig:oneloop_newbasis}, respectively (diagrams in the same row have the same spurion structure).
Note that each term in \cref{eq:L_dipole} must be free of logarithms like $\log (\mat{m}_R^\dagger\mat{m}_R^{})$. 
UV logarithms are trivially absent because we started from a renormalizable theory and the field redefinition does not introduce a counterterm for the dipole operator. 
Meanwhile, the absence of IR logarithms can be understood from either the IR-finiteness of the diagrams, or from the absence of tree-level-generated operators that can renormalize the dipole at one loop. 
We can also see that $U(1)_\text{EM}^{}$ charge spurions for the fermions, which we commented on below \cref{eq:PLR_Q} but did not write out explicitly, are not relevant here: for the charged components of $L$, $L^c$ to enter the diagrams in the second row of \cref{fig:oneloop_newbasis}, we would need both Higgs fields at the $\mat{c}_{6L}$ vertex to be neutral, contradicting the fact that $\mat{c}_{6L}$ (at its physical value) couples the charged lepton $l^2=e$ to a charged Higgs $\Sigma\indices{^1_2} = H^+$. 
Finally, it is clear from the diagrams that each term in the expression of $\mathcal{O}_\text{dipole}$ can only involve one fermion mass; therefore, additional spurion combinations that one may naively write down consistent with the requirements above, {\it e.g.}\ $\bigl[ \mat{m}_R^{-1\dagger} \bigr]\indices{_{\Ri{r'}}^{\Li{t'}}} \bigl[ \mat{m}_R^{} \bigr]\indices{^{\Ri{u'}}_{\Li{t'}}} \bigl[ \mat{m}_R^\dagger \bigr]\indices{^{\Li{s'}}_{\Ri{u'}}}$ in place of $\bigl[ \mat{m}_R^{} \bigr]\indices{_{\Ri{r'}}^{\Li{s'}}}$ in the first term of \cref{eq:L_dipole} and its $P_{LR}^{}$ counterpart, must be forbidden because they yield terms that involve both $m_S^{}$ and $m_E^{}$.

We can now substitute in expressions for $\mat{c}_{5L}^{}$, $\mat{c}_{5R}^{}$, $\mat{c}_{6L}^{}$, $\mat{c}_{6R}^{}$ from \cref{eq:c5,eq:c6L,eq:c6R}. 
The two terms in \cref{eq:L_dipole} can be combined by relabelling indices and noting that $F^{\alpha\beta}$ is symmetric while fermion fields are anti-commuting (which means $F^{\alpha\beta}\, r^c_{\alpha\li{i'}}\, l_\beta^{\li{j}} = -F^{\beta\alpha}\, l_\beta^{\li{j}}\, r^c_{\alpha\li{i'}}$). 
We can further simplify the result using the antisymmetry of $U(1)_\text{EM}^{}$ charge tensor, $Q\Hi{\indices{^m_{n'}^p_{q'}}} = -Q\Hi{\indices{^p_{q'}^m_{n'}}}$.
For the terms proportional to $Y_V^{\prime*}$, we find
\begin{align}
\mathcal{O}_\text{dipole}^{(Y_V')} \propto &\;
\bigl[\mat{m}_L^{} \mat{\Delta m}^{-2}\, \mat{m}_L^{-1} -\mat{m}_R^{-1} \mat{\Delta \bar m}^{-2} \,\mat{m}_R^{}\bigr]\Li{\indices{_u^{\,v}_{\,r'}^{s'}}}
\nonumber\\
&\quad \times \bigl[\mat{y}_L^{}\bigr]\indices{_{\li{i}}^{\,\Li{r'}}_{\Hi{p}}^{\,\Hi{q'}}} 
\bigl[\mat{y}_V^{\prime\dagger}\bigr]\indices{_{\Li{s'}}^{\Li{u}}_{\,\Hi{k}}^{\,\Hi{l'}}}
\bigl[\mat{y}_R^{}\bigr]\indices{_{\Li{v}}^{\,\li{j'}}_{\Hi{m}}^{\,\Hi{n'}}}
\,Q\Hi{\indices{^m_{n'}^p_{q'}}}
\,e\,F^{\alpha\beta}\,l^{\li{i}}_\alpha\,r^c_{\beta\li{j'}}\,\Sigma\Hi{\indices{^k_{l'}}} 
\;\;\to\;\; 0\,.
\end{align}
This vanishes because the tensor $\bigl[\mat{m}_L^{} \mat{\Delta m}^{-2}\, \mat{m}_L^{-1} -\mat{m}_R^{-1} \mat{\Delta \bar m}^{-2} \,\mat{m}_R^{}\bigr]$ is identically zero at the physical values of mass spurions, \cref{eq:phys_m}.
For the terms proportional to $Y_V$, we find:
\begin{align}
\mathcal{O}_\text{dipole}^{(Y_V)} \propto &\; \biggl\{
\bigl[\mat{m}_R^{-1}\bigr]\indices{_{\Li{r'}}^{\Ri{s'}}} \bigl[\mat{m}_L^{-1}\bigr]\indices{_{\Ri{u}}^{\,\Li{v}}}
\,\Bigl( Q\Hi{\indices{^k_{l'}^p_{q'}}} \,\Sigma\Hi{\indices{^m_{n'}}} -Q\Hi{\indices{^k_{l'}^m_{n'}}} \,\Sigma\Hi{\indices{^p_{q'}}}\Bigr) \nonumber\\
&\qquad 
+\bigl[ \mat{m}_L^{-1}\, \mat{m}_R^{-1}\, \mat{\Delta \bar m}^{-2}\, \mat{m}_R^{}\, \mat{m}_R^\dagger -\mat{m}_L^\dagger \mat{m}_L^{}\, \mat{\Delta m}^{-2}\, \mat{m}_L^{-1}\, \mat{m}_R^{-1}\bigr]\indices{_{\Ri{u}}^{\,\Li{v}}_{\,\Li{r'}}^{\Ri{s'}}} \,Q\Hi{\indices{^m_{n'}^p_{q'}}} \,\Sigma\Hi{\indices{^k_{l'}}}
\biggr\} \nonumber\\
&\quad \times \bigl[\mat{y}_L^{}\bigr]\indices{_{\li{i}}^{\,\Li{r'}}_{\Hi{p}}^{\,\Hi{q'}}} \bigl[\mat{y}_V^{}\bigr]\indices{_{\Ri{s'}}^{\Ri{u}}_{\,\Hi{k}}^{\,\Hi{l'}}} \bigl[\mat{y}_R^{}\bigr]\indices{_{\Li{v}}^{\,\li{j'}}_{\Hi{m}}^{\,\Hi{n'}}} 
\,e\, F^{\alpha\beta}\,l^{\li{i}}_\alpha\,r^c_{\beta\li{j'}} \,.
\end{align}
At the physical values of mass spurions, the expression in curly brackets,
\begin{equation}
\biggl\{\dots\biggr\} \to \delta_u^v \,\delta_{r'}^{s'} \,\frac{1}{m_L\, m_{R,s'}}\, \Bigl( Q\indices{^k_{l'}^p_{q'}} \,\Sigma\indices{^m_{n'}} -Q\indices{^k_{l'}^m_{n'}} \,\Sigma\indices{^p_{q'}} +Q\indices{^m_{n'}^p_{q'}} \,\Sigma\indices{^k_{l'}}\Bigr) \,,
\end{equation}
where $m_{R,s'}$ are the diagonal elements of $\mat{m}_R^{}$, {\it i.e.}\ $m_{R,1} = m_S$, $m_{R,2} = m_E (\to \infty)$. 
Meanwhile, as far as the $SU(2)_L^{}$ indices are concerned, the product of Yukawa spurions is proportional to
\begin{equation}
\bigl[\mat{y}_L^{}\bigr]\indices{_i^{\,r'}_p^{\,q'}} \bigl[\mat{y}_V^{}\bigr]\indices{_{s'}^u_{\,k}^{\,l'}} \bigl[\mat{y}_R^{}\bigr]\indices{_v^{\,j'}_m^{\,n'}}  \propto \epsilon_{ip}\,\delta^u_k\,\epsilon_{vm}\,,
\end{equation}
which is simply the statement that $SU(2)_L$ is unbroken at the physical values of these spurions. 
Finally, for the Yukawa spurions to be nonzero, we must pick $q'=2$, $l'=n'=1$ (see \cref{eq:phys_y}) --- this reflects the fact that $Y_L^{}$ couples to $H$ whereas $Y_R^{}$ and $Y_V^{}$ couple to $H^*$. 
We therefore obtain
\begin{equation}
\mathcal{O}_\text{dipole}^{(Y_V)} \propto \epsilon_{km}\,\Bigl( Q\indices{^k_{1}^p_{q'}} \,\Sigma\indices{^m_{1}} -Q\indices{^k_{1}^m_{1}} \,\Sigma\indices{^p_{q'}} +Q\indices{^m_{1}^p_{q'}} \,\Sigma\indices{^k_{1}}\Bigr) 
\;\;\to\;\;0 \,,
\end{equation}
noting that the first and third terms in the parentheses are symmetric in $k$ and $m$, while the second term vanishes because $Q\indices{^k_{1}^m_{1}} =0$. 
(In fact, we did not even need the value of $q'$; setting $l'=n'$ was sufficient to show $\mathcal{O}_\text{dipole}^{(Y_V)}$ vanishes.) 
We have therefore proved, via a spurion analysis in the $SU(2)_L^{}\times SU(2)_R^{} \times P_{LR}^{}$ symmetric theory, that the dimension-six dipole operator must have a vanishing coefficient at one-loop level in the model of \cref{eq:model}. 
The proof also extends to certain higher-loop contributions where propagators and/or vertices in the diagrams in \cref{fig:oneloop_newbasis} are dressed without changing the $SU(2)_L^{}\times SU(2)_R^{}$ symmetry structure.

It is worth reflecting on the role of field redefinition in this proof. 
Normally, when constructing an EFT operator basis, one performs field redefinitions to eliminate higher-dimensional operators in favor of lower-dimensional ones. 
In the case of SMEFT, for example, we start with an operator basis up to dimension four, {\it i.e.}\ the renormalizable SM.
Next, when enumerating operators at dimension six, we use field redefinitions (equivalently, equations of motion) to eliminate as many dimension-six operators as possible in favor of other dimension-six operators plus lower-dimensional operators in the renormalizable SM, in order to arrive at a nonredundant basis~\cite{Grzadkowski:2010es}.
The same procedure can be carried out through higher operator dimensions. 
In this section, however, we used field redefinitions in the opposite direction --- to eliminate lower-dimensional operators (dimension-four Yukawa) in favor of higher-dimensional ones (dimension-five and six operators in \cref{eq:L_dim5,eq:L_dim6} in particular).
Crucially, higher-dimensional operators come with more fields and/or more derivatives. 
This significantly restricts the number of possibilities when enumerating spurion combinations for an operator generated at certain loop order that has a limited number of derivatives ({\it e.g.}\ the dipole operator at one loop, with no derivatives acting on the Higgs). 

\section{UV-IR conspiracy?}
\label{sec:conspiracy}

We have seen that the magic zero in vector-like singlet and doublet fermion contributions to the muon's dipole moments can be understood from the perspective of symmetries. However, that still leaves what might be perceived as a UV-IR conspiracy from the perspective of an effective field theorist living between the mass scales of the singlet and doublet fermions when $m_S \ll m_L$. As detailed in Ref.~\cite{Arkani-Hamed:2021xlp}, such an observer would see an interaction of the form
\begin{equation} \label{eq:irtree}
\mathcal{L}_{\rm eff} \supset \frac{Y_R^{} Y_V^{\prime*}}{m_L^2}\,S^{c \dag} H^\dag \not \!\! D (H^\dag e^c) \,,
\end{equation}
and be tempted to estimate the muon dipole moments by computing a diagram where $S^c$ and $H$ are closed into a loop with a $Y_L$ Yukawa, and an external photon is attached. Presumably the effective field theorist would attempt the calculation in a mass-independent scheme, in which case they would obtain a finite ``infrared'' result. This could only be reconciled with the measured vanishing of the dipole moments by invoking exact cancellation of this ``infrared'' piece with another ``ultraviolet'' piece that come from one-loop matching at the scale $m_L$. From the perspective of an effective field theorist between $m_S$ and $m_L$ it may look like a violation of Wilsonian naturalness. The field redefinition we used in \cref{sec:redef} to disentangle contributions at $m_L$ and $m_S$ only sharpens the problem.

As a practical matter, perhaps the effective field theorist would not be so confused. The scales $m_S$ and $m_L$ would be separated by at most a loop factor in the presence of nonzero $Y_V^{}, Y'_V$ because of RG mixing, and observing the interaction in \cref{eq:irtree}, together with other tree- and loop-level irrelevant operators, would give a clear picture of the physics at the scale $m_L$. Setting these practical considerations aside, though, the apparent conspiracy is a fascinating phenomenon that merits further consideration.

As hinted above, however, the apparent conspiracy is scheme dependent. For example, one can use Pauli-Villars regularization when computing loop diagrams. For the IR contribution coming from the $S$ fermion exchange diagram with the operator in \cref{eq:irtree}, this introduces a factor $M_{PV}^{2}/(M_{PV}^{2}-k^{2})$ where $M_{PV}^{}$ is to be taken to infinity at the end of the calculation. Then the $k$ integral becomes zero, as the procedure effectively makes $h(0)=0$ in the notation of \cref{sec:infrared}. Similarly, the UV matching contribution also vanishes in this scheme. Of course, the introduction of regulators does not influence the final amplitude for the dipole, but it does change the apparent distribution between UV and IR contributions. The fact that the IR contribution obtained via Pauli-Villars regularization matches the full amplitude is specific to this case -- and would not be true in a more general example -- but clearly illustrates the role of scheme dependence.

The effective field theorist in this scenario would have a strong incentive to perform the calculation in dimensional regularization with a mass-independent renormalization scheme, since that would be guaranteed to preserve EFT power counting while Pauli-Villars would not. But the fact that the partition between UV and IR contributions is scheme dependent is a helpful reminder not to ascribe too much significance to apparent conspiracies in a given scheme. Indeed, this is already quite familiar to aficionados of the hierarchy problem: the problem itself only becomes completely well-defined when the Higgs mass is rendered finite and various contributions scheme independent. 

That said, our hapless effective field theorist working in dimensional regularization would face an apparent UV-IR conspiracy eerily reminiscent of hierarchy problems we face in the present era, making it sensible to speculate about whether this situation might arise in other contexts. In this particular case, we encountered an apparent UV-IR conspiracy in which the observable in question was zero, but similar situations could arise when the observable is IR-dominated (even if nonzero). Such IR dominance would seem to be a property of irrelevant operators, rather than relevant ones. This makes it less obvious how analogous UV-IR conspiracies might explain the scale problems of the Standard Model, but it would be delightful to be proven wrong.

\section{Conclusions}
\label{sec:conclusions}

In this paper we considered one-loop contributions to the muon's electric and magnetic dipole moments from new vector-like singlet and doublet leptons. The vanishing of these contributions at leading nontrivial order -- an instance of a ``magic zero'' -- has been characterized as a calculable violation of the Wilsonian notion of naturalness. Here we showed, however, that the magic zero can be explained by symmetries, preserving (to our knowledge) the sanctity of Wilsonian naturalness in calculable, non-gravitational, local, Lorentz-invariant field theories in four dimensions. Our explanation involves a spurion analysis grounded in the symmetries of the free theory, which must be augmented by power-counting and established non-renormalization theorems. Crucially, the observed zero can be understood without constructing or evaluating any loop integrals. 

The zero leads to an apparent UV-IR conspiracy at intermediate scales. While the apparent conspiracy's scheme dependence is a helpful reminder not to ascribe undue physical significance to how the calculation is organized, it nonetheless illustrates the surprises that can arise when partitioning contributions to an observable in an EFT framework. If nothing else, it underlines the care that should be taken in invoking Wilsonian naturalness when irrelevant operators are involved. 

There may well be other explanations for this particular magic zero, grounded in symmetry or not. Invocation of infrared dominance and/or total derivative phenomena provides a valid explanation for vanishing loop diagrams starting from the form of the integrand, and reveals interesting relationships between operators that would otherwise remain obscure. As always, the success of any explanation can be judged by its predictiveness. The techniques employed in our analysis may be helpful in constructing or understanding additional magic zeroes, which may in turn shed new light on long-standing naturalness problems of the Standard Model and beyond. Even in more pedestrian contexts, magic zeroes akin to the one studied here may play a useful role, protecting order-of-magnitude separations among otherwise comparable rates or scales. At the very least, our results suggest that reports of the death of Wilsonian naturalness are greatly exaggerated.

\acknowledgments
We thank S.~Knapen for helpful conversations, and are indebted to N.~Arkani-Hamed for helpful conversations and comments on the manuscript.
The work of IGG and AV is supported by the National Science Foundation under Grant No.~NSF PHY-1748958, as well as by the Gordon and Betty Moore Foundation through Grant GBMF7392 (IGG). The work of NC and ZZ is supported by the U.S.~Department of Energy under the grant DE-SC0011702. NC thanks LBNL and the BCTP for hospitality during the completion of this work.

\appendix
\section{Appendix}
\label{sec:app}

\subsection{Cancellation of $Y_V^{}$ terms}
\label{sec:app_calculation}

\begin{figure}[h]
	\centering
	\includegraphics[scale=1]{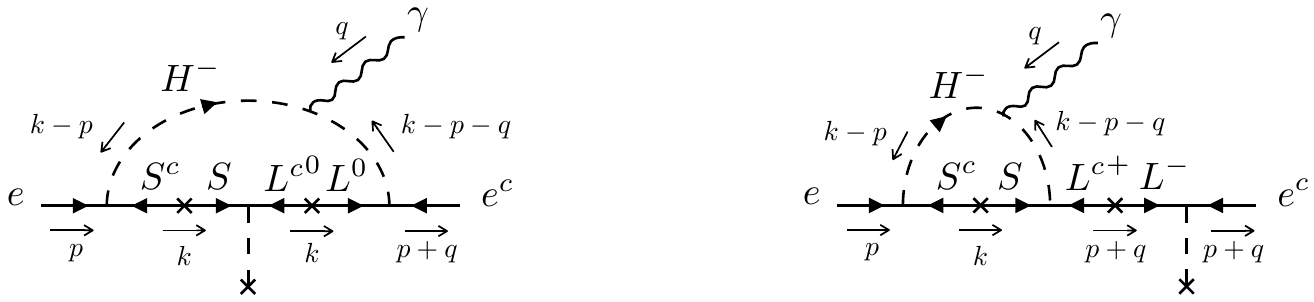}
	\caption{One-loop diagrams providing the (a priori) leading contribution to the electromagnetic dipole moment of charged SM leptons in the model of \cref{eq:model} with $Y'_V = 0$. Both diagrams are proportional to the combination of couplings $Y_L^{} Y_R^{} Y_V^{}$.}
	\label{fig:oneloop_YV}
\end{figure}

In the model of \cref{eq:model}, there are two one-loop diagrams contributing to the dipole moment of the SM fermions that are proportional to $Y_V^{}$. These are shown in \cref{fig:oneloop_YV}.
Ignoring overall multiplicative vertex factors, the amplitude of the left diagram in \cref{fig:oneloop_YV} is proportional to
\begin{align}
i \mathcal{M}^\mu_\text{(left)} 
& \propto \int \frac{d^4 k}{(2 \pi)^4} \frac{m_L m_S \, (2 k^\mu - 2 p^\mu - q^\mu)}{(k-p)^2 (k-p-q)^2 (k^2 - m_L^2) (k^2 - m_S^2)} \bar u (p+q) P_L u(p) , 
\end{align}
whereas for the one on the right we have:
\begin{align}
i \mathcal{M}^\mu_\text{(right)} 
& \propto \int \frac{d^4 k}{(2 \pi)^4} \frac{- m_L m_S \, (2 k^\mu - 2 p^\mu - q^\mu)}{(k-p)^2 (k-p-q)^2 ((p+q)^2 - m_L^2) (k^2 - m_S^2)} \bar u (p+q) P_L u(p) . 
\end{align}
In total, adding the contribution from both diagrams and expanding to linear order in external momenta we find:
\begin{align}
& i \mathcal{M}^\mu_\text{(left)} + i \mathcal{M}^\mu_\text{(right)} \propto \nonumber \\
& \qquad \frac{m_S}{m_L} \int \frac{d^4 k}{(2 \pi)^4} \frac{\bar u (p+q) P_L u(p)}{k^2 (k^2 - m_L^2) (k^2 - m_S^2)} \left\{ - (2 p^\mu + q^\mu) + 2 k^\mu \left( 1 + \frac{2 k \cdot (2 p + q)}{k^2} \right) \right\} .
\end{align}
This is identical to \cref{eq:Mmu_linear}, except for the overall factor of $m_S / m_L$. Following the discussion in \cref{sec:1loop}, the integrand of the loop function vanishses at linear order in $p$ and $q$, and therefore the dipole contribution at this order is zero.

\subsection{Mass basis analysis of the $Y_V^{}$ terms}
\label{sec:app_broken}

The mass basis calculation of the $Y_V^{}$ terms proceeds very much like the discussion in \cref{sec:mass}. Setting $Y'_V = 0$ and keeping terms propotional to $Y_V$, the masses and Yukawa couplings of the toy model of \cref{sec:toy} remain as given in \cref{eq:fulltotoy1,eq:fulltotoy2}, except for the expression for $\hat y_R^*$, which now reads
\begin{equation}
{\hat y}_R^* = Y_R^{}\, \theta^* - Y_V^{}\, \theta_+ \,.
\end{equation}
Here $\theta_\pm$ refer to mixing angles in the charged fermion sector, defined in terms of the flavor-to-mass basis rotation as:
\begin{equation}
\begin{pmatrix} {L^c}^+ \\ e^c \end{pmatrix} \rightarrow \begin{pmatrix} 1 & \theta_+ \\ - \theta_+^* & 1 \end{pmatrix} \begin{pmatrix} {L^c}^+ \\ e^c \end{pmatrix} \qquad \text{and} \qquad
\begin{pmatrix} L^- \\ e \end{pmatrix} \rightarrow \begin{pmatrix} 1 & \theta_- \\ - \theta_-^* & 1 \end{pmatrix} \begin{pmatrix} L^- \\ e \end{pmatrix} .
\end{equation}
To $\mathcal{O} ( v / m_L)$ and $\mathcal{O} (v / m_S)$ the mixing angles are given by
\begin{align}
\theta = \frac{v}{\sqrt{2}} \frac{m_L^{} Y_V^*}{|m_L|^2 - |m_S|^2}, \qquad \theta^c = \frac{v}{\sqrt{2}} \frac{m_S^{} Y_V^*}{|m_L|^2 - |m_S|^2} , \qquad \theta_+ = \frac{v}{\sqrt{2}}\,\frac{Y_R^{}}{m_L} ,
\end{align}
and $\theta_- = 0$. Up to overall numerical factors, the contribution to the electromagnetic dipole moment from each individual massive fermion satisfies
\begin{equation}
e \,\frac{y_L y_R^*}{m} = - e \,\frac{{\hat y}_L {\hat y}_R^*}{\hat m} = - \frac{v}{\sqrt{2}} \frac{m_S^*}{m_L} \frac{e \, Y_L Y_R Y_V}{|m_L|^2 - |m_S|^2} ,
\end{equation}
and the combination in \cref{eq:delta_d} indeed vanishes.

In this case, the vanishing of the leading contribution to the dipole term comes down to the mixing angles in the fermion sector satisfying the following relationship:
\begin{equation}
{\theta^c}^* = \frac{m_L}{m_S} \left( \theta^* - \theta_+ \frac{Y_V}{Y_R} \right) .
\label{eq:magic_YV}
\end{equation}
Proceeding as in \cref{sec:mass}, before rotating to the mass basis, the mass terms in the neutral fermion sector read
\begin{align}
\mathcal{L} \supset	& \left\{ - m_L L^0 {L^c}^0 - m_S S S^c - \frac{Y_V v}{\sqrt{2}} {L^c}^0 S \right\} + \text{h.c.} 
\label{eq:Lmass_0}
\end{align}
This is identical to \cref{eq:L_flavor}, after the substitutions $L^0 \leftrightarrow {L^c}^0$, $S \leftrightarrow S^c$, as well as $Y'_V \rightarrow Y_V$. The mixing angles are therefore those in \cref{eq:mixingangles} after further exchanging $\theta^c \leftrightarrow \theta^*$. In total:
\begin{equation}
\theta = N \frac{m_L Y_V^* v /\sqrt{2}}{|m_L|^2 - |m_S|^2} \qquad \text{and} \qquad
\theta^c = N \frac{m_S Y_V^* v /\sqrt{2}}{|m_L|^2 - |m_S|^2} .
\end{equation}
Additionally, the mass terms in the charged fermion sector are given by
\begin{equation}
\mathcal{L} \supset \left\{ m_L L^- {L^c}^+ - \frac{Y_R v}{\sqrt{2}} L^- e^c \right\} + \text{h.c.}
\end{equation}
This is identical to \cref{eq:Lmass_0} after making the substitutions $\{ {L^c}^0, L^0 \} \rightarrow \{ L^-, {L^c}^+ \}$ and $\{ S^c, S \} \rightarrow \{ e, e^c \}$, together with $m_L \rightarrow - m_L$, $m_S \rightarrow 0$, and $Y_V \rightarrow Y_R$. The mixing angles in the charged sector are therefore given by
\begin{equation}
\theta_+ = - \left. \theta^* \right|_{\{ \star \}}  \qquad \text{and} \qquad \theta_- = - \left. \theta^c \right|_{\{ \star \}} ,
\end{equation}
with $\{ \star \} \equiv \{ m_L \rightarrow - m_L, m_S \rightarrow 0, Y_V \rightarrow Y_R \}$. Explicitly:
\begin{equation}
\theta_+ = N \frac{Y_R v /\sqrt{2}}{m_L} \qquad \text{and} \qquad \theta_- = 0 .
\end{equation}
One can readily verify that \cref{eq:magic_YV} is indeed satisfied by all these mixing angles.

\subsection{Additional details of field redefinition}
\label{sec:app_redef}

In this appendix, we supplement the discussion in \cref{sec:redef} with additional details. 
The goal is to find the unitary matrices $\Mat{O}$, $\Mat{O}^c$ that solve \cref{eq:redef_diag}. 
Unitarity of these matrices, $\Mat{O}^\dagger\Mat{O} = \Mat{O}^{c\dagger}\Mat{O}^c = \Mat{1}$, implies $\Mat{O}_{(1)} +\Mat{O}_{(1)}^\dagger = 0$, $\Mat{O}_{(2)} +\Mat{O}_{(2)}^\dagger = \bigl(\Mat{O}_{(1)}\bigr)^2$, etc., and likewise for $\Mat{O}^c$.
The forms of $\Mat{O}_{(n)}$, $\Mat{O}^c_{(n)}$ are further constrained by 
the fact that the Higgs field $\Sigma$ carries one $SU(2)_L^{}$ index and one $SU(2)_R^{}$ index. 
Upon imposing these constraints, we have, up to the second order:
\begin{align}
&
\Mat{O}_{(1)}^{} = 
\begin{pmatrix}
0 & 0 & \mat{O}_{lR}^{} \\
0 & 0 & \mat{O}_{LR}^{} \\
-\mat{O}^\dagger_{lR} & -\mat{O}^\dagger_{LR} & 0
\end{pmatrix} 
\,,\qquad
&
\Mat{O}_{(2)}^{} = 
\begin{pmatrix}
\cdot & \mat{O}_{lL}^{} & 0 \\
\mat{O}_{Ll}^{} & \cdot & 0 \\
0 & 0 & \;\;\cdot\;\;
\end{pmatrix} 
\,,\label{eq:O1O2}\\[6pt]
&
\Mat{O}^c_{(1)} =
\begin{pmatrix}
0 & 0 & \mat{O}^c_{rL} \\
0 & 0 & \mat{O}^c_{RL} \\
-\mat{O}^{c\dagger}_{rL} & -\mat{O}^{c\dagger}_{RL} & 0
\end{pmatrix} 
\,,\qquad
&
\Mat{O}^c_{(2)} = 
\begin{pmatrix}
\cdot & \mat{O}^c_{rR} & 0 \\
\mat{O}^c_{Rr} & \cdot & 0 \\
0 & 0 & \;\;\cdot\;\;
\end{pmatrix} 
\,,\label{eq:Oc1Oc2}
\end{align}
where ``$\cdot$'' denotes a nonzero entry whose detailed form will not be needed. 
The nonzero entries written out above are further related by $\mat{O}_{Ll}^{} +\mat{O}^\dagger_{lL} +\mat{O}_{LR}^{}\mat{O}^\dagger_{lR} = \mat{O}^c_{Rr} + \mat{O}^{c\dagger}_{rR} +\mat{O}^c_{RL}\mat{O}^{c\dagger}_{rL} =0$. 

The condition \cref{eq:redef_diag} can be equivalently written as
\begin{equation}
\begin{cases}
\Mat{O}^\dagger\, (\Mat{M}+\Mat{U}) \, (\Mat{M}+\Mat{U})^\dagger \,\Mat{O} = 
\text{diag} \Bigl( 
\mathcal{O}\bigl((\mat{y}\Sigma)^2\bigr) \,,\; 
\mat{m}_L^{} \mat{m}_L^\dagger + \mathcal{O}\bigl((\mat{y}\Sigma)^2\bigr) \,,\; 
\mat{m}_R^{} \mat{m}_R^\dagger + \mathcal{O}\bigl((\mat{y}\Sigma)^2\bigr) 
\Bigr)
\,, \\[8pt]
\Mat{O}^{c\dagger}\, (\Mat{M}+\Mat{U})^\dagger  (\Mat{M}+\Mat{U}) \,\Mat{O}^c = 
\text{diag} \Bigl( 
\mathcal{O}\bigl((\mat{y}\Sigma)^2\bigr) \,,\; 
\mat{m}_R^\dagger \mat{m}_R^{} + \mathcal{O}\bigl((\mat{y}\Sigma)^2\bigr) \,,\; 
\mat{m}_L^\dagger \mat{m}_L^{} + \mathcal{O}\bigl((\mat{y}\Sigma)^2\bigr) 
\Bigr) \,.
\end{cases}
\label{eq:diag_unbroken}
\end{equation}
Note the absence of $\mathcal{O}(\mat{y}\Sigma)$ terms on the RHS of \cref{eq:diag_unbroken} due to $SU(2)_L^{}\times SU(2)_R^{}$ representation mismatch.
Expanding these equations in $\mat{y}\Sigma$, we obtain, up to the second order:
\begin{align}
\mathcal{O}(\mat{y}\Sigma):& \qquad 
\begin{cases}
\;\bigl[ \Mat{O}_{(1)}^{} \,,\; \Mat{M} \,\Mat{M}^\dagger \bigr] \;=\; \Mat{M}\,\Mat{U}^\dagger +\Mat{U}\,\Mat{M}^\dagger \,,\\
\;\bigl[ \Mat{O}_{(1)}^c \,,\; \Mat{M}^\dagger \,\Mat{M} \bigr] \;=\; \Mat{M}^\dagger\Mat{U} +\Mat{U}^\dagger\Mat{M} \,,
\end{cases}
\label{eq:cond_1}\\
\mathcal{O}\bigl((\mat{y}\Sigma)^2\bigr):& \qquad 
\begin{cases}
\Mat{O}_{(2)}^{} \Mat{M}\,\Mat{M}^\dagger + \Mat{M}\,\Mat{M}^\dagger \Mat{O}_{(2)}^\dagger = \Mat{U} \,\Mat{U}^\dagger + \Mat{O}_{(1)}^{} \Mat{M}\,\Mat{M}^\dagger \Mat{O}_{(1)}^{} +\text{diag}(\,\cdot\,,\,\cdot\,,\,\cdot\,) \,,\\
\Mat{O}_{(2)}^c \Mat{M}^\dagger\Mat{M} + \Mat{M}^\dagger\Mat{M}\, \Mat{O}_{(2)}^{c\dagger} = \Mat{U}^\dagger\Mat{U} + \Mat{O}_{(1)}^c \Mat{M}^\dagger\Mat{M}\, \Mat{O}_{(1)}^c +\text{diag}(\,\cdot\,,\,\cdot\,,\,\cdot\,) \,,
\end{cases}
\label{eq:cond_2}
\end{align}
where we have used \cref{eq:cond_1} and the relations $\Mat{O}_{(2)} +\Mat{O}_{(2)}^\dagger = \bigl(\Mat{O}_{(1)}\bigr)^2$, $\Mat{O}_{(2)}^c +\Mat{O}_{(2)}^{c\dagger} = \bigl(\Mat{O}_{(1)}^c\bigr)^2$ to simplify the $\mathcal{O}\bigl((\mat{y}\Sigma)^2\bigr)$ terms to derive \cref{eq:cond_2}.
Solving the first order equations, \cref{eq:cond_1}, we find
\begin{align}
\begin{cases}
\mat{O}_{lR}^{} = \mat{U}_{lR}^{} \,\mat{m}_R^{-1} \,, \\
\mat{O}^c_{rL} = \mat{U}_{Lr}^\dagger\, \mat{m}_L^{-1\dagger} \,,
\end{cases} 
\qquad\quad
\begin{cases}
\bigl[\mat{O}_{LR}^{}\bigr]\indices{_i^{\,j'}} = \bigl[\mat{\Delta\bar{m}}^{-2}\bigr]\indices{_i^{\,k}_{\,l'}^{j'}} \,\bigl[ \mat{U}_{LR}^{} \mat{m}_R^\dagger + \mat{m}_L^{} \mat{U}_{RL}^\dagger\bigr]\indices{_k^{\,l'}} \,,\\[2pt]
\bigl[\mat{O}^c_{RL}\bigr]\indices{_{i'}^j} = -\bigl[\mat{\Delta m}^{-2}\bigr]\indices{_l^{\,j}_{\,i'}^{k'}}\, \bigl[ \mat{U}_{LR}^\dagger \mat{m}_L^{} +\mat{m}_R^\dagger\mat{U}_{RL}^{}\bigr]\indices{_{k'}^l} \,,
\end{cases}
\label{eq:O_sol_1}
\end{align}
where $\mat{\Delta\bar{m}}^{-2}$, $\mat{\Delta m}^{-2}$ are the inverse of the mass squared difference tensors defined in \cref{eq:dm2,eq:dmbar2}.
Solving the second order equations, \cref{eq:cond_2}, we find
\begin{equation}
\begin{cases}
\mat{O}_{Ll}^{} = \mat{m}_L^{-1\dagger}\, \mat{U}_{RL}^\dagger \, \mat{m}_R^{-1\dagger}\, \mat{U}_{lR}^\dagger \,,\\
\mat{O}^c_{Rr} = \mat{m}_R^{-1} \,\mat{U}_{RL}^{} \,\mat{m}_L^{-1} \,\mat{U}_{Lr}^{} \,,
\end{cases}
\qquad\quad
\begin{cases}
\mat{O}_{lL}^{} = -\mat{U}_{lR}^{}\, \mat{m}_R^{-1}\,\bigl( \mat{U}_{RL}^{}\,\mat{m}_L^{-1} +\mat{O}_{LR}^\dagger\bigr) \,,\\
\mat{O}_{rR}^c = -\mat{U}_{Lr}^\dagger \mat{m}_L^{-1\dagger} \bigl(\mat{U}_{RL}^\dagger \mat{m}_R^{-1\dagger} +\mat{O}_{RL}^{c\dagger} \bigr)\,.
\end{cases}
\label{eq:O_sol_2}
\end{equation}
As in the main text,  we leave $SU(2)_L^{}\times SU(2)_R^{}$ indices implicit when their contraction can be unambiguously inferred from matrix multiplication. 
The field redefinition preserves $P_{LR}^{}$ provided the latter acts on the rotation matrices as ${\Mat{O}^*}\indices{^I_J} \leftrightarrow {\Mat{O}^c}\indices{_I^{\,J}}$. 
One can readily verify that the equations grouped together in \cref{eq:O_sol_1,eq:O_sol_2} are related by $P_{LR}^{}$.

\subsection{Gauge basis spurion analysis for the Yukawa operator}
\label{sec:app_yukawa}

We can carry out a similar spurion analysis as in \cref{sec:explain} for the dimension-four Yukawa operator and see explicitly that, in contrast to the dimension-six dipole, it is not required by symmetries to vanish. 
We will also see that the spurion analysis reproduces the correct mass dependence of muon Yukawa threshold correction.

To construct the Yukawa operator, we impose the same requirements listed in \cref{sec:explain}, with the following exceptions:
\begin{itemize}
	\item We should replace $e\,F^{\alpha\beta}\,Q\Hi{\indices{^m_{n'}^p_{q'}}} \to \epsilon^{\alpha\beta} \,\epsilon^{\Hi{mp}}\,\epsilon_{\Hi{n'q'}}$.
	\item The operator coefficient should have mass dimension zero instead of $-2$, and logarithms can appear. 
\end{itemize}
The Yukawa operator is then constrained to take the following unique form:
\begin{align}
\mathcal{O}_\text{Yukawa} \propto &\;
l^{\alpha\li{i}}\, \bigl[\mat{c}_{5L}^{}\bigr]\indices{_{\li{i}}^{\,\Ri{r'}}_{\Hi{p}}^{\,\Hi{q'}}} \bigl[ \mat{m}_R^{} \mat{m}_R^\dagger \mat{m}_R^{} \,f(\mat{m}_R^\dagger \mat{m}_R^{}) \bigr]\indices{_{\Ri{r'}}^{\Li{s'}}} \bigl[\mat{c}_{6R}^{}\bigr]\indices{_{\Li{s'}}^{\li{j'}}_{\Hi{k}}^{\,\Hi{l'}}_{\Hi{m}}^{\,\Hi{n'}}}  r^c_{\alpha\li{j'}}\,\Sigma\Hi{\indices{^k_{l'}}}\,\epsilon^{\Hi{mp}}\,\epsilon_{\Hi{n'q'}} \nonumber\\
&+
r^{c\,\alpha}_{\li{i'}}\, \bigl[\mat{c}_{5R}^{}\bigr]\indices{_{\Ri{r}}^{\,\li{i'}}_{\Hi{q}}^{\,\Hi{p'}}} \bigl[f(\mat{m}_L^{} \mat{m}_L^\dagger) \,\mat{m}_L^{} \mat{m}_L^\dagger \mat{m}_L^{} \bigr]\indices{_{\Li{s}}^{\,\Ri{r}}} \bigl[\mat{c}_{6L}^{}\bigr]\indices{_{\li{j}}^{\Li{\,s}}_{\Hi{l}}^{\,\Hi{k'}}_{\Hi{n}}^{\,\Hi{m'}}}  l_\alpha^{\li{j}}\,\Sigma\Hi{\indices{^l_{k'}}}\,\epsilon^{\Hi{nq}}\,\epsilon_{\Hi{m'p'}} \,,
\label{eq:L_yukawa}
\end{align}
where
\begin{equation}
f(x) \equiv \alpha \log\bigl(x/\mu^2\bigr) + \beta \,,
\end{equation}
with $\alpha$, $\beta$ constant and $\mu$ the renormalization scale. 
We can combine both terms in \cref{eq:L_yukawa}, noting that $r^{c\,\alpha}_{\li{i'}}\,l_\alpha^{\li{j}} = l^{\alpha\li{j}}\,r^{c}_{\alpha\li{i'}}$ (without extra minus sign, in contrast to the dipole case). 
Substituting in expressions for $\mat{c}_{5L}^{}$, $\mat{c}_{5R}^{}$, $\mat{c}_{6L}^{}$, $\mat{c}_{6R}^{}$ from \cref{eq:c5,eq:c6L,eq:c6R}, we find, for the terms proportional to $Y_V^{\prime*}$:
\begin{align}
\mathcal{O}_\text{Yukawa}^{(Y_V')} \propto &\;
\bigl[f(\mat{m}_R^\dagger \mat{m}_R^{})\, \mat{m}_R^\dagger \mat{m}_R^{}\, \mat{m}_L^{} \mat{\Delta m}^{-2}\, \mat{m}_L^{-1} -\mat{m}_R^{-1} \mat{\Delta \bar m}^{-2} \,\mat{m}_R^{}\, \mat{m}_L^{}\mat{m}_L^\dagger\, f(\mat{m}_L^{}\mat{m}_L^\dagger)\bigr]\Li{\indices{_u^{\,v}_{\,r'}^{s'}}}
\nonumber\\
&\quad \times \bigl[\mat{y}_L^{}\bigr]\indices{_{\li{i}}^{\,\Li{r'}}_{\Hi{p}}^{\,\Hi{q'}}} 
\bigl[\mat{y}_V^{\prime\dagger}\bigr]\indices{_{\Li{s'}}^{\Li{u}}_{\,\Hi{k}}^{\,\Hi{l'}}}
\bigl[\mat{y}_R^{}\bigr]\indices{_{\Li{v}}^{\,\li{j'}}_{\Hi{m}}^{\,\Hi{n'}}}
\,\epsilon^{\Hi{mp}}\,\epsilon_{\Hi{n'q'}}
\,l^{\alpha\li{i}}\,r^c_{\alpha\li{j'}}\,\Sigma\Hi{\indices{^k_{l'}}} \,.
\end{align}
As opposed to $\bigl[\mat{m}_L^{} \mat{\Delta m}^{-2}\, \mat{m}_L^{-1} -\mat{m}_R^{-1} \mat{\Delta \bar m}^{-2} \,\mat{m}_R^{}\bigr]=0$ in the dipole case, we now have
\begin{align}
&\bigl[f(\mat{m}_R^\dagger \mat{m}_R^{})\, \mat{m}_R^\dagger \mat{m}_R^{}\, \mat{m}_L^{} \mat{\Delta m}^{-2}\, \mat{m}_L^{-1} -\mat{m}_R^{-1} \mat{\Delta \bar m}^{-2} \,\mat{m}_R^{}\, \mat{m}_L^{}\mat{m}_L^\dagger\, f(\mat{m}_L^{}\mat{m}_L^\dagger)\bigr]\indices{_u^{\,v}_{\,r'}^{s'}} \nonumber\\[4pt]
=&\; \delta_u^v\,\delta_{r'}^{s'} \,\Biggl[\frac{\alpha}{|m_{R,s'}^{}|^2 -|m_L^{}|^2} \, \biggl(|m_{R,s'}^{}|^2 \log\frac{|m_{R,s'}^{}|^2}{\mu^2} -|m_L^{}|^2\log\frac{|m_L^{}|^2}{\mu^2} \biggr) +\beta\Biggr] \,.
\label{eq:L_yukawa_1}
\end{align}
For the terms proportional to $Y_V$, we find:
\begin{align}
\mathcal{O}_\text{Yukawa}^{(Y_V)} \propto &\; \biggl\{
\bigl[ f(\mat{m}_R^\dagger \mat{m}_R^{}) \,\mat{m}_R^\dagger\bigr]\indices{_{\Li{r'}}^{\Ri{s'}}} \bigl[\mat{m}_L^{-1}\bigr]\indices{_{\Ri{u}}^{\,\Li{v}}}
\, \epsilon^{\Hi{kp}}\,\epsilon_{\Hi{l'q'}}\,\Sigma\Hi{\indices{^m_{n'}}} \nonumber\\
&\quad +\bigl[ \mat{m}_R^{-1}\bigr]\indices{_{\Li{r'}}^{\Ri{s'}}} \bigl[\mat{m}_L^\dagger\,f(\mat{m}_L^{}\mat{m}_L^\dagger)\bigr]\indices{_{\Ri{u}}^{\,\Li{v}}}
\,\epsilon^{\Hi{km}}\,\epsilon_{\Hi{l'n'}} \,\Sigma\Hi{\indices{^p_{q'}}} 
\nonumber\\
&\quad 
+\bigl[ \mat{m}_L^{-1}\, \mat{m}_R^{-1}\, \mat{\Delta \bar m}^{-2}\, \mat{m}_R^{}\, \mat{m}_R^\dagger \mat{m}_L^{}\, \mat{m}_L^\dagger \,f(\mat{m}_L^{}\, \mat{m}_L^\dagger ) \nonumber\\
&\qquad\; - f(\mat{m}_R^\dagger \mat{m}_R^{}) \,\mat{m}_R^\dagger \mat{m}_R^{}\,\mat{m}_L^\dagger \mat{m}_L^{}\, \mat{\Delta m}^{-2}\, \mat{m}_L^{-1}\, \mat{m}_R^{-1}\bigr]\indices{_{\Ri{u}}^{\,\Li{v}}_{\,\Li{r'}}^{\Ri{s'}}} \,\epsilon^{\Hi{mp}}\,\epsilon_{\Hi{n'q'}} \,\Sigma\Hi{\indices{^k_{l'}}}
\biggr\} \nonumber\\
&\quad \times \bigl[\mat{y}_L^{}\bigr]\indices{_{\li{i}}^{\,\Li{r'}}_{\Hi{p}}^{\,\Hi{q'}}} \bigl[\mat{y}_V^{}\bigr]\indices{_{\Ri{s'}}^{\Ri{u}}_{\,\Hi{k}}^{\,\Hi{l'}}} \bigl[\mat{y}_R^{}\bigr]\indices{_{\Li{v}}^{\,\li{j'}}_{\Hi{m}}^{\,\Hi{n'}}} 
\,l^{\alpha\li{i}}\,r^c_{\alpha\li{j'}} \,.
\end{align}
At the physical values of mass spurions, the expression in curly brackets,
\begin{align}
\biggl\{\dots\biggr\} \to \delta_u^v \,\delta_{r'}^{s'} \,\Biggl[
&
\frac{m_{R,s'}^*}{m_L^{}}\,f\bigl(|m_{R,s'}^{}|^2\bigr) 
\biggl( \epsilon^{kp}\,\epsilon_{l'q'} \,\Sigma\indices{^m_{n'}} -\frac{|m_L^{}|^2}{|m_{R,s'}^{}|^2 -|m_L^{}|^2} \,\epsilon^{mp}\,\epsilon_{n'q'}\,\Sigma\indices{^k_{l'}} \biggr) \nonumber\\
&
+\frac{m_L^*}{m_{R,s'}^{}}\, f\bigl( |m_L^{}|^2\bigr) 
\biggl( \epsilon^{km}\,\epsilon_{l'n'}\,\Sigma\indices{^p_{q'}} +\frac{|m_{R,s'}^{}|^2}{|m_{R,s'}^{}|^2-|m_L^{}|^2} \,\epsilon^{mp}\,\epsilon_{n'q'}\,\Sigma\indices{^k_{l'}} \biggr)
\Biggr] \,.
\end{align}
As discussed in \cref{sec:explain}, the product of Yukawa spurions is proportional to $\epsilon_{km}$ and is nonzero only for $l'=n'=1$. 
We therefore obtain
\begin{align}
\mathcal{O}_\text{Yukawa}^{(Y_V)} \propto &\; \frac{|m_{R,s'}^{}|^2}{|m_{R,s'}^{}|^2-|m_L^{}|^2}\,
\biggl[
\frac{m_{R,s'}^*}{m_L^{}}\,f\bigl(|m_{R,s'}^{}|^2\bigr) 
-\frac{m_L^*}{m_{R,s'}^{}}\, f\bigl( |m_L^{}|^2\bigr) 
\biggr]\, \epsilon_{1q'}\,\Sigma\indices{^p_1} \nonumber\\
=&\;
\frac{m_{R,s'}^*}{m_L^{}} \,\Biggl[
\frac{\alpha}{|m_{R,s'}^{}|^2-|m_L^{}|^2} \,
\biggl(|m_{R,s'}^{}|^2 \log\frac{|m_{R,s'}^{}|^2}{\mu^2} -|m_L^{}|^2\log\frac{|m_L^{}|^2}{\mu^2}\biggr)
+\beta
\Biggr] \,.
\label{eq:L_yukawa_2}
\end{align}
Finally, for both $\mathcal{O}_\text{Yukawa}^{(Y_V')}$ ($\mathcal{O}_\text{Yukawa}^{(Y_V)}$), the Yukawa coupling spurion $\mat{y}_V^{\prime\dagger}$ ($\mat{y}_V^{}$) picks out the $s'=1$ tensor component, so $m_{R,s'}^{}$ in the equations above are set to $m_{R,1}^{}=m_S^{}$. 
We see that \cref{eq:L_yukawa_1,eq:L_yukawa_2} reproduce the correct mass dependence of the muon Yukawa threshold correction.

\bibliography{magiczeros_refs}
	
\end{document}